\documentclass{article}
\usepackage{graphicx}
\usepackage{amsmath,amssymb,amsfonts}

\begin{document}
\bibliographystyle{unsrt}

\title{Effective-Field-Theory Approach to \\ \medskip Top-Quark Production and Decay}
\author{{\bf Cen Zhang and Scott Willenbrock}\\ \\
Department of Physics, University of Illinois at Urbana-Champaign \\ 1110 West Green Street, Urbana, IL  61801}
\maketitle
\begin{abstract}
We discuss new physics in top-quark interactions, using an effective field theory approach. We consider top-quark decay, single top production, and top-quark pair production.  We identify 15 dimension-six operators that contribute to these processes, and we compute the deviation from the Standard Model induced by these operators. The results provide a systematic way of searching for (or obtaining bounds on) physics beyond the Standard Model.
\end{abstract}

\section{Introduction}
The standard model (SM) of the strong and electroweak interactions has been very successful phenomenologically. However, there are reasons to believe that the SM is not a complete description of nature. It contains many arbitrary parameters with no connections, and provides no explanation for the symmetry-breaking mechanism that gives rise to the masses of gauge bosons and fermions. In addition, the existence of dark matter and dark energy are not accommodated by the SM. It also provides no explanation of the strong CP problem. Consequently, there could be further particles and interactions as one probes higher energy scales. These energy scales could be the Planck scale or an intermediate scale $\Lambda$.

A complete description of any new physics beyond the SM requires a fundamental theory. There are many models proposed for new physics. For example, the new particles could be those of supersymmetric theories or fermions that transform under different representations of the SM gauge group. The new interactions may also originate from quark substructure or preon exchange. Because of the diversity of these models, it is useful to introduce a model-independent approach.

If the physics beyond the SM lies at an energy scale $\Lambda$ less than 1 TeV, then we should be able to observe it directly at high-energy colliders. If it lies at a scale much greater than 1 TeV, then we can parameterize its effects via higher-dimension operators, suppressed by inverse powers of the scale $\Lambda$. If the new physics is too heavy to appear directly in low energy processes, then we can integrate it out from the Lagrangian. In this way, the effective Lagrangian is:
\begin{equation}
L_\mathit{eff}=L_0+\frac{1}{\Lambda}L_1+\frac{1}{\Lambda^2}L_2+\cdots
\end{equation}
where $L_0$ is the SM Lagrangian of dimension four, $L_1$ is the new interaction of dimension five, $L_2$ is of dimension six, etc. The procedure is quite general and independent of the new interactions at scale $\Lambda$. The only constraint is that all $L_i$ are ${\rm SU(3)}_C\times {\rm SU(2)}_L\times {\rm U(1)}_Y$ invariant.

At dimension five, the only operator allowed by gauge invariance is \cite{Weinberg:1979sa}
\begin{equation}
L_\mathit{eff}=\frac{c^{ij}}{\Lambda}(L^{iT}\epsilon\phi)C(\phi^T\epsilon L^j)+h.c.
\end{equation}
where $L^i$ is the lepton doublet field of the $i^{th}$ generation and $\phi$ is the Higgs doublet field. When the Higgs doublet acquires a vacuum-expectation value, this term gives rise to a Majorana mass for neutrinos. Due to the tiny neutrino masses, the scale $\Lambda$ is probably around $10^{15}$ GeV. In contrast to this unique dimension-five operator, there are many independent dimension-six operators \cite{Leung:1984ni,Buchmuller:1985jz}.

Because the top quark is heavy relative to all the other observed SM fermions, we expect that the new physics at higher energy scales may reveal itself at lower energies through the effective interactions of the top quark, and deviations with respect to the SM predictions might be detectable. In this paper we will study the effect of these dimension-six operators on top quark interactions at hadron colliders. We focus on three different processes: top quark decay, single top production, and top pair production. If no deviation is observed experimentally, then one can place bounds on the coefficients of the dimension-six operators. The effects of non-standard interactions on top-quark physics at linear colliders and photon colliders can be found in Refs.~{\cite{linear1,linear2,4fermi,photon}}. 

We choose the effective Lagrangian to realize the weak symmetry linearly, as the precision electroweak data favors a light Higgs boson. The situation where the weak symmetry is realized nonlinearly is studied in Ref.~\cite{Tait:2000sh,Peccei:1989kr,Peccei:1990uv,Larios:1996ib,Malkawi:1994tg}. We will use the operator set introduced by Buchmuller and Wyler \cite{Buchmuller:1985jz}. In their paper they categorize all possible gauge-invariant dimension-six operators, and use the equations of motion (EOM) to simplify them into 80 independent operators (for one generation).  Subsequently it was found that several of these operators are actually not independent \cite{Grzadkowski:2003tf,AguilarSaavedra:2009mx,Grzadkowski:2010es}. We focus on the operators that have an influence on the top quark.

We expect the leading modification to SM processes at order $\frac{1}{\Lambda^2}$. In this paper we don't consider higher order contributions. We expect the scale $\Lambda$ to be large (at least larger than the scale we can probe directly) so $\frac{1}{\Lambda^4}$ contributions would be small compared to the uncertainty on top quark measurements. Thus we ignore all dimension-eight (and higher) operators, as well as effects involving two dimension-six operators.

For any physical observable, the $\frac{1}{\Lambda^2}$ contribution comes from the interference between dimension-six operators and the SM Lagrangian. This contribution might be suppressed for a variety reasons. For example, since all quark and lepton masses are negligible compared to the top quark mass, a new interaction that involves a right-handed quark or lepton (except for the top quark) has a very small interference with the SM charged-current weak interactions, which only involve left-handed fermions.
It turns out that although there are a large number of dimension-six operators, only a few of them have significant effects at order $\frac{1}{\Lambda^2}$. We list these operators in Tables 1 and 2.
\begin{table}[htb]
\centering
\begin{tabular}{|l|l|}
\hline
operator & process
\\\hline
$O_{\phi q}^{(3)}=i(\phi^+\tau^ID_\mu\phi)(\bar{q}\gamma^\mu\tau^Iq)$ & top decay, single top
\\\hline
\raisebox{-.5ex}{$O_{tW}=(\bar{q}\sigma^{\mu\nu}\tau^It)\tilde{\phi}W^I_{\mu\nu}$ (with real coefficient)} & \raisebox{-.5ex}{top decay, single top}
\\\hline
$O_{qq}^{(1,3)}=(\bar{q}^i\gamma_\mu\tau^I q^j)(\bar{q}\gamma^\mu\tau^I q)$ & single top
\\\hline
\raisebox{-.5ex}{$O_{tG}=(\bar{q}\sigma^{\mu\nu}\lambda^A t)\tilde{\phi}G^A_{\mu\nu}$ (with real coefficient)} & \raisebox{-.5ex}{single top, $q\bar{q},gg\rightarrow t\bar{t}$}
\\\hline
$O_{G}=f_{ABC}G^{A\nu}_\mu G^{B\rho}_\nu G^{C\mu}_\rho$ & $gg\rightarrow t\bar{t}$
\\\hline
$O_{\phi G}=\frac{1}{2}(\phi^+\phi)G^A_{\mu\nu}G^{A\mu\nu}$ & $gg\rightarrow t\bar{t}$
\\\hline
7 {\rm four-quark operators} & $q\bar q\rightarrow t\bar{t}$
\\\hline
\end{tabular}
\caption{CP-even operators that have effects on top-quark processes at order $1/\Lambda^2$. Here $q$ is the left-handed quark doublet, while $t$ is the right-handed top quark. The field $\phi$ ($\tilde{\phi}=\epsilon\phi^*$) is the Higgs boson doublet. $D_\mu=\partial_\mu-ig_s\frac{1}{2}\lambda^AG_\mu^A-ig\frac{1}{2}\tau^IW_\mu^I-ig'YB_\mu$ is the covariant derivative. $W_{\mu\nu}^I=\partial_\mu W_\nu^I-\partial_\nu W^I_\mu+g\epsilon_{IJK}W^J_\mu W^K_\nu$ is the $W$ boson field strength, and $G_{\mu\nu}^A=\partial_\mu G_\nu^A-\partial_\nu G^A_\mu+g_sf^{ABC}G^B_\mu G^C_\nu$ is the gluon field strength. Because of the Hermiticity of the Lagrangian, the coefficients of these operators are real, except for $O_{tW}$ and $O_{tG}$. The operator $O_{\phi q}^{(3)}$ with an imaginary coefficient can be removed using the EOM.}
\end{table}
\begin{table}[htb]
\centering
\begin{tabular}{|l|l|}
\hline
operator & process
\\\hline
\raisebox{-.5ex}{$O_{tW}=(\bar{q}\sigma^{\mu\nu}\tau^It)\tilde{\phi}W^I_{\mu\nu}$ (with imaginary coefficient)} & \raisebox{-.5ex}{top decay, single top}
\\\hline
\raisebox{-.5ex}{$O_{tG}=(\bar{q}\sigma^{\mu\nu}\lambda^A t)\tilde{\phi}G^A_{\mu\nu}$ (with imaginary coefficient)} & \raisebox{-.5ex}{single top, $q\bar{q},gg\rightarrow t\bar{t}$}
\\\hline
\raisebox{-.5ex}{$O_{\tilde{G}}=f_{ABC}\tilde{G}^{A\nu}_\mu G^{B\rho}_\nu G^{C\mu}_\rho$} & \raisebox{-.5ex}{$gg\rightarrow t\bar{t}$}
\\\hline
\raisebox{-.5ex}{$O_{\phi \tilde{G}}=\frac{1}{2}(\phi^+\phi)\tilde{G}^A_{\mu\nu}G^{A\mu\nu}$} & \raisebox{-.5ex}{$gg\rightarrow t\bar{t}$}
\\\hline
\end{tabular}
\caption{CP-odd operators that have effects on top-quark processes at order $1/\Lambda^2$. Notations are the same as in Table 1, and $\tilde{G}_{\mu\nu}=\epsilon_{\mu\nu\rho\sigma}G^{\rho\sigma}$.}
\end{table}

In Table 1, only one of the four-quark operators, $O_{qq}^{(1,3)}=(\bar{q}^i\gamma_\mu\tau^I q^j)(\bar{q}\gamma^\mu\tau^I q)$, is listed explicitly. Here the superscripts $i,j$ denote the first two quark generations, while $q$ without superscript denotes the third generation. In single top production, this is the only (independent) four-quark operator that contributes. However, there are many other four-quark operators with different isospin and color structures \cite{Leung:1984ni,Buchmuller:1985jz}. In the top pair production process $q\bar{q}\rightarrow t\bar{t}$, seven such operators contribute. The details are discussed in Section 4.

In Table 2, the CP-odd operators are listed. These interactions interfere with the SM only if the spin of the top quark is taken into account.  The reason is that the SM conserves CP to a good approximation (the only CP violation is in the CKM matrix), and the inteference between a CP-odd operator and a CP-even operator is a CP violation effect. It was shown in Ref.~\cite{Atwood:2000tu} that, in the absence of final-state interactions, any CP violation observable can assume non-zero value only if it is $T_N$-odd, where $T_N$ is the ``naive'' time reversal, which means to apply time reversal without interchanging the initial and final states. Thus an observable is $T_N$-odd if it is proportional to a term of the form $\epsilon_{\mu\nu\rho\sigma}v^\mu v^\nu v^\rho v^\sigma$. If we don't consider the top quark spin, $v$ must be the momentum of the particles, and such a term will not be present because the reactions we consider here involve at most three independent momenta. Therefore top polarimetry is essential for the study of CP violation.
Since the top quark rapidly undergoes two-body weak decay $t\rightarrow Wb$ with a time much shorter than the time scale necessary to depolarize the spin, information on the top spin can be obtained from its decay products. CP violation will be discussed in Section 5.

There is an argument that can be used to neglect some of the new operators \cite{Arzt:1994gp}. Some new operators can be generated at tree level from an underlying gauge theory, while others must be generated at loop order. In general the loop generated operators are suppressed by a factor of $1/16\pi^2$. However, the underlying theory may not be a weakly coupled gauge theory, or the loop diagrams could be enhanced due to the index of a fermion in a large representation. Furthermore, the underlying theory may not be a gauge theory at all. Fortunately, the effective field theory approach does not depend on the underlying theory. We will consider all dimension-six operators, without making any assumptions about the nature of the underlying theory.

We do not make any assumptions about the flavor structure of the
dimension-six operators, although we don't consider any flavor-changing
neutral currents in this paper.  The charged-current weak interaction of
the top quark is proportional to $V_{tb}$, so the SM rate for top decay
and single top production is proportional to $V_{tb}^2$.  We write
all dimension-six operators in terms of mass-eigenstate fields, so no
diagonalization of the new interactions is necessary. Hence, in
charged-current weak interactions, the interference
between the SM amplitude and the new interaction is proportional to
$V_{tb}C_i$, where $C_i$ is the (real) coefficient of the dimension-six Hermitian
operator $O_i$ (also recall that $V_{tb}$ itself is purely real in the standard
parameterization \cite{Amsler:2008zzb}). If the operator is not Hermitian, the coefficient $C_i$ is complex;
CP-conserving processes are proportional to $V_{tb}{\rm Re}C_i$, while CP-violating processes are instead proportional to $V_{tb}{\rm Im}C_i$.

Deviations of top-quark processes from SM predictions have often been
discussed using a vertex-function approach, where the $Wtb$ vertex is
parameterized in terms of four unknown form factors \cite{Kane:1991bg}.  Given
our precision knowledge of the electroweak interaction, this approach is
too crude.  The effective field theory approach is well motivated; it
takes into consideration the unbroken ${\rm SU(3)}_C\times {\rm
SU(2)}_L\times {\rm U(1)}_Y$ gauge symmetry; it includes contact
interactions as well as vertex corrections; it is valid for both
on-shell and off-shell quarks; and it can be used for loop processes
\cite{Gomis:1995jp}. None of these virtues are shared by the vertex function
approach \cite{Zhang:2010px}.

\section{Top Quark Decay}

When the fermion masses (except for the top quark) are ignored, there are only two independent dimension-six operators in \cite{Buchmuller:1985jz} that contribute to top-quark decay at leading order:\footnote{The operator $O_{Dt}=(\bar{q}D_\mu t)D^\mu\tilde{\phi}$ listed in Ref.~\cite{Buchmuller:1985jz} can be removed using the EOM \cite{AguilarSaavedra:2009mx}.}
\begin{eqnarray}
O_{\phi q}^{(3)}&=&i(\phi^+\tau^ID_\mu\phi)(\bar{q}\gamma^\mu\tau^Iq)\\
O_{tW}&=&(\bar{q}\sigma^{\mu\nu}\tau^It)\tilde{\phi}W^I_{\mu\nu}
\end{eqnarray}
The operators $O_{\phi q}^{(3)}$ and $O_{tW}$ modify the SM $Wtb$ interaction.
Upon symmetry breaking, they generate the following terms in the Lagrangian:
\begin{eqnarray}
&&L_\mathit{eff}=\frac{C_{\phi q}^{(3)}}{\Lambda^2}\frac{gv^2}{\sqrt{2}}\bar{b}\gamma^\mu P_LtW^-_\mu+h.c.\\
&&L_\mathit{eff}=-2\frac{C_{tW}}{\Lambda^2}v\bar{b}\sigma^{\mu\nu}P_Rt \partial_\nu W^-_\mu+h.c.
\end{eqnarray}
where $v=246$ GeV is the vacuum expectation value (VEV) of $\phi$. The operator $O_{\phi q}^{(3)}$ simply leads to a rescaling of the SM $Wtb$ vertex by a factor of $(1+\frac{C_{\phi q}^{(3)}v^2}{\Lambda^2V_{tb}})$, so it does not affect any distributions, and is therefore impossible to detect in angular distributions of top-quark decays.  The vertex-function approach to top-quark decay is pursued in Refs.~\cite{AguilarSaavedra:2006fy,AguilarSaavedra:2007rs}.

\begin{figure}[htb]
\centering\includegraphics[width=12cm]{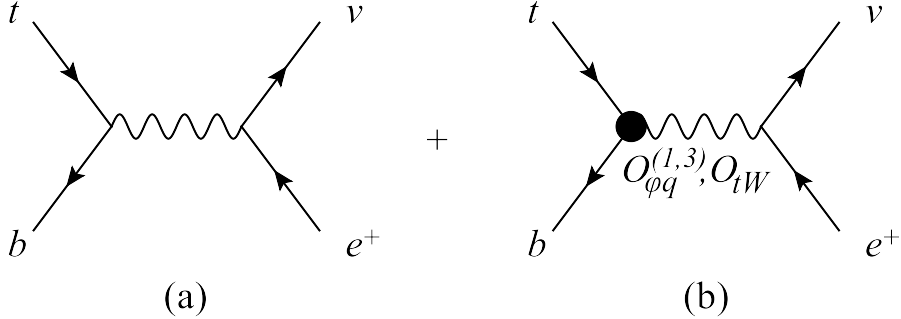}
\caption{
The Feynman diagrams for $t\rightarrow be^+\nu$. (a) is the SM amplitude; (b) represents the vertex correction induced by the operator $O_{\phi q}^{(3)}$ and $O_{tW}$.}\label{fig1}
\end{figure}

These operators interfere with the SM amplitude, as is shown in Figure \ref{fig1}. We can compute their correction to the SM amplitude. The $t\rightarrow be^+\nu$ squared amplitude is:
\begin{equation}
\frac{1}{2}\Sigma|M|^2=\frac{V_{tb}^2g^4u(m_t^2-u)}{2(s-m_W^2)^2}+\frac{C_{\phi q}^{(3)}V_{tb}v^2}{\Lambda^2}\frac{g^4u(m_t^2-u)}{(s-m_W^2)^2}+\frac{4\sqrt{2}{\rm Re}C_{tW}V_{tb}m_tm_W}{\Lambda^2}\frac{g^2su}{(s-m_W^2)^2}
\end{equation}
where $C_i$ is the coefficient of operator $O_i$, and $s,t,u$ are generalizations of the usual Mandelstam variables ($s=(p_t-p_b)^2, \quad t=(p_t-p_\nu)^2, \quad u=(p_t-p_{e^+})^2$). $C_{\phi q}^{(3)}$ is real.

Using the narrow width approximation for the $W$ boson, the differential decay rate is
\begin{eqnarray}
\frac{d\Gamma}{d\cos\theta}&=&\left(V_{tb}^2+\frac{2C_{\phi q}^{(3)}V_{tb}v^2}{\Lambda^2}\right)\frac{g^4}{4096\pi^2m_t^3m_W\Gamma_W}
(m_t^2-m_W^2)^2[m_t^2+m_W^2+(m_t^2-m_W^2)\cos\theta](1-\cos\theta)\nonumber\\
&&+\frac{{\rm Re}C_{tW}V_{tb}g^2}{128\sqrt{2}\pi^2\Lambda^2m_t^2\Gamma_W}m_W^2(m_t^2-m_W^2)^2(1-\cos\theta)
\end{eqnarray}
Here $\theta$ is the angle between the momenta of top quark and the neutrino in the $W$ rest frame, and $\Gamma_W$ is the width of the $W$ boson. In the SM, at tree level $\Gamma_W$ is given by:
\begin{equation}
\Gamma_W=\frac{3\alpha_W}{4}m_W\;.
\end{equation}

\begin{figure}[t]
\centering\includegraphics[width=8cm]{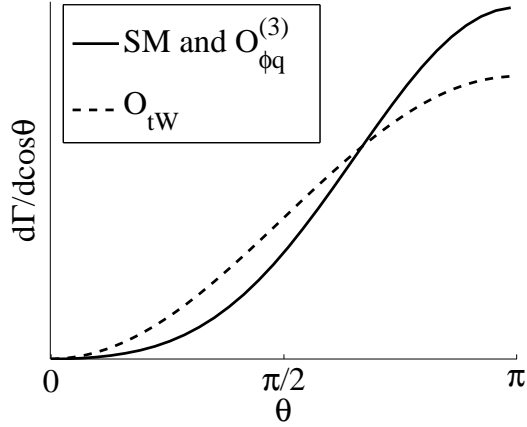}
\caption{The differential decay rate induced by different operators. The curves are normalized so that the area is the same.}\label{fig2}
\end{figure}

The angular dependence is shown in Figure \ref{fig2}. The curves are normalized to have equal areas. The contribution from $O_{\phi q}^{(3)}$ is the same as the SM contribution, because $O_{\phi q}^{(3)}$ simply rescales the SM $Wtb$ vertex. It therefore does not affect angular distributions. The angular dependence of the contribution from $O_{tW}$ is not dramatically different from the SM.

The partial width is given by
\begin{equation}
\Gamma=\left(V_{tb}^2+\frac{2C_{\phi q}^{(3)}V_{tb}v^2}{\Lambda^2}\right)\frac{g^4(m_t^6-3m_W^4m_t^2+2m_W^6)}{3072\pi^2\Gamma_Wm_t^3m_W}
+{\rm Re}C_{tW}V_{tb}\frac{g^2m_W^2(m_t^2-m_W^2)^2}{64\sqrt{2}\pi^2\Lambda^2\Gamma_Wm_t^2}\;.
\end{equation}
Both dimension-six operators affect the partial width.  The total width is given by the above expression times a factor of nine.  Unfortunately, it is not known how to measure the partial or total widths in a hadron collider environment.

\begin{figure}[t]
\begin{minipage}[c]{8cm}
\centering\includegraphics[width=8cm]{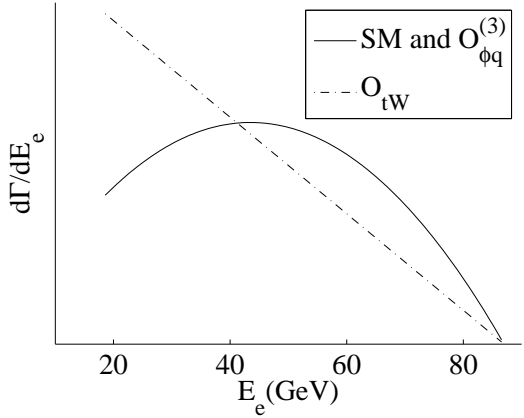}
\caption{The energy dependence of the electron.}\label{fig3}
\end{minipage}
\hspace{0cm}
\begin{minipage}[c]{8cm}
\centering\includegraphics[width=8cm]{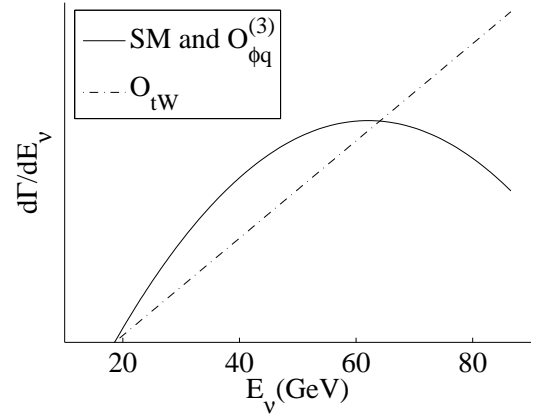}
\caption{The energy dependence of the neutrino.}\label{fig4}
\end{minipage}
\end{figure}

We also consider the energy dependence of the leptons in the top quark rest frame. The SM computation can be found in \cite{Jezabek:1994zv,Jezabek:1994qs}. The correction from dimension-six operators at leading order is:
\begin{eqnarray}
\frac{d\Gamma}{dE_{e^+}}&=&\left(V_{tb}^2+\frac{2C_{\phi q}^{(3)}V_{tb}v^2}{\Lambda^2}\right)
\frac{g^4E_{e^+}(m_t-2E_{e^+})}{128\pi^2m_W\Gamma_W}+\frac{{\rm Re}C_{tW}V_{tb}g^2m_W^2(m_t-2E_{e^+})}{16\sqrt{2}\pi^2\Lambda^2\Gamma_W}
\nonumber\\
\frac{d\Gamma}{dE_\nu}&=&
\left(V_{tb}^2+\frac{2C_{\phi q}^{(3)}V_{tb}v^2}{\Lambda^2}\right)
\frac{g^4(-4E_\nu^2m_t^2+2E_\nu(m_t^3+2m_W^2m_t)-m_W^2(m_t^2+m_W^2))}{256\pi^2m_t^2m_W\Gamma_W}\nonumber\\
&&+
\frac{{\rm Re}C_{tW}V_{tb}g^2m_W^2(2E_\nu m_t-m_W^2)}{16\sqrt{2}\pi^2\Lambda^2m_t\Gamma_W}
\end{eqnarray}
where $m_W^2/2m_t<E_{e^+},E_\nu<m_t/2$ are the energies of the electron and neutrino, respectively. We don't list the energy dependence of the bottom quark, because the narrow width approximation for the $W$ boson is used and the energy of the bottom quark is given by $E_b=(m_t^2-m_W^2)/2m_t$. These results are shown in Figure \ref{fig3} and \ref{fig4}. Again the curves are normalized so that the areas are the same. Compared to Figure \ref{fig2}, the two curves are more distinct, which implies the effect of $O_{tW}$ would be more apparent in the energy distribution of the leptons.

The angular distribution and the energy distribution are not independent. The energy of the leptons are fixed in the $W$ rest frame. Therefore their energy in the top quark rest frame is given by a boost, which only depends on the angle $\theta$:
\begin{eqnarray}
E_\nu=\frac{1}{2}(E+|q|\cos\theta)\\
E_{e^+}=\frac{1}{2}(E-|q|\cos\theta)
\end{eqnarray}
where $E=(m_t^2+m_W^2)/2m_t$ and $|q|=(m_t^2-m_W^2)/2m_t$ are the energy and momentum of the $W$ boson in the top quark rest frame. Furthermore, both the angular distribution and energy distribution can be expressed using the $W$ helicity fractions \cite{AguilarSaavedra:2006fy}:
\begin{equation}
\frac{1}{\Gamma}\frac{d\Gamma}{d\cos\theta}=
\frac{3}{8}(1+\cos\theta)^2F_R
+\frac{3}{8}(1-\cos\theta)^2F_L
+\frac{3}{4}\sin^2\theta F_0
\end{equation}
and
\begin{equation}
\frac{1}{\Gamma}\frac{d\Gamma}{dE_{e^+}}=\frac{1}{(E_{max}-E_{min})^3}
\left(
3(E_{e^+}-E_{min})^2F_R+3(E_{max}-E_{e^+})^2F_L+6(E_{max}-E_{e^+})(E_{e^+}-E_{min})F_0
\right)
\end{equation}
where $E_{max}=m_t/2$ and $E_{min}=m_W^2/2m_t$, $F_i=\Gamma_i/\Gamma$ are the $W$ boson helicity fractions, corresponding to positive (R), negative (L), or zero (0) helicity. The helicity fraction is affected by
the operator $O_{tW}$:
\begin{eqnarray}
F_0&=&\frac{m_t^2}{m_t^2+2m_W^2}-\frac{4\sqrt{2}{\rm Re}C_{tW}v^2}{\Lambda^2V_{tb}}\frac{m_tm_W(m_t^2-m_W^2)}{(m_t^2+2m_W^2)^2}\nonumber\\
F_L&=&\frac{2m_W^2}{m_t^2+2m_W^2}+\frac{4\sqrt{2}{\rm Re}C_{tW}v^2}{\Lambda^2V_{tb}}\frac{m_tm_W(m_t^2-m_W^2)}{(m_t^2+2m_W^2)^2}\nonumber\\
F_R&=&0
\end{eqnarray}
These equations make manifest the earlier observation that the operator $O_{\phi q}^{(3)}$, which simply rescales the SM vertex, cannot affect any distributions.  Thus top-quark decay is sensitive only to the operator $O_{tW}$, and can be used to measure (or bound) its coefficient.

Finally, we investigate the polarized differential decay rate. In the rest frame of the top quark, the angular distribution of any top quark decay product is given by \cite{Jezabek:1994zv,Jezabek:1994qs}
\begin{equation}\label{define_alpha}
\frac{1}{\Gamma}\frac{d\Gamma}{d\cos\theta_i}=\frac{1+\alpha_i\cos\theta_i}{2}
\end{equation}
where $\theta_i=\theta_b, \theta_v, \theta_{e^+}$ is the angle between the spin axis of the top quark and the momentum of the bottom quark, neutrino or positron. The ``analyzing power'' $\alpha_i$ measures the degree to which the direction of the decay product $i$ is correlated with the top spin. If dimension-six operators are added, the relation still holds, but the coefficient $\alpha_i$ will be affected by the new operators. Since $O_{\phi q}^{(3)}$ is just a rescaling of the SM interaction, the only correction is from $O_{tW}$. This could be an independent way to determine the coefficient ${\rm Re}C_{tW}$. At leading order, the correction is given by:
\begin{eqnarray}
\alpha_b&=&-\frac{m_t^2-2m_W^2}{m_t^2+2m_W^2}
+\frac{{\rm Re}C_{tW}v^2}{\Lambda^2V_{tb}}
\frac{8\sqrt{2}m_tm_W(m_t^2-m_W^2)}{(m_t^2+2m_W^2)^2}\nonumber\\
\alpha_v&=&\frac{m_t^6-12m_t^4m_W^2+3m_t^2m_W^4(3+8\ln(m_t/m_W))+2m_W^6}{m_t^6-3m_t^2m_W^4+2m_W^6}\nonumber\\
&&
-\frac{{\rm Re}C_{tW}v^2}{\Lambda^2V_{tb}}
\frac{12\sqrt{2}m_tm_W(m_t^6-6m_t^4m_W^2+3m_t^2m_W^4(1+4\ln(m_t/m_W))+2m_W^6)}{(m_t^2+2m_W^2)^2(m_t^2-m_W^2)^2}\nonumber\\
\alpha_{e^+}&=&1\label{eq:anapower1}
\end{eqnarray}
The same equations hold for hadronic top decay, with $\alpha_u=\alpha_\nu$, $\alpha_{\bar d}=\alpha_{e^+}$.
The coefficient $\alpha_{e^+}$ is not affected by dimension-six operators. This is consistent with the results in Ref.~\cite{Grzadkowski:2002gt}.

The measurement of these coefficients requires a source of polarized top quarks. This is addressed in the next section.

\section{Single Top Production}

Single top quarks are produced through the electroweak interaction. There are three separate processes: $s$-channel \cite{Cortese:1991fw}, $t$-channel \cite{Willenbrock:1986cr,Yuan:1989tc,Ellis:1992yw}, and $Wt$ production \cite{Heinson:1996zm}. An effective field theory approach to the $s$- and $t$-channel processes was advocated in Ref.~\cite{Cao:2007ea}.  We update that analysis by including an additional operator, which was neglected in that study because it is loop-suppressed if the underlying theory is a gauge theory.  We also perform an effective field theory analysis of the $Wt$ process.  The vertex-function approach to single-top production is pursued in Refs.~\cite{Boos:1999dd,Chen:2005vr,AguilarSaavedra:2008gt}.

Single top production contains four distinct channels: the $s$-channel process $u\bar{d}\rightarrow t\bar{b}$, the $t$-channel processes $ub\rightarrow dt$ and $\bar{d}b\rightarrow \bar{u}t$, and the $Wt$ associated production channel $gb\rightarrow Wt$. We first consider the $s$ and $t$ channels.
The following operators contribute \cite{Cao:2007ea}:
\begin{eqnarray}
O_{\phi q}^{(3)}&=&i(\phi^+\tau^ID_\mu\phi)(\bar{q}\gamma^\mu\tau^Iq)\\
O_{tW}&=&(\bar{q}\sigma^{\mu\nu}\tau^It)\tilde{\phi}W^I_{\mu\nu}\\
O_{qq}^{(1,3)}&=&(\bar{q}^i\gamma_\mu\tau^I q^j)(\bar{q}\gamma^\mu\tau^I q)
\end{eqnarray}
For the four-quark operator $O_{qq}^{(1,3)}$, the superscripts $i,j$ denote the first two quark generations. Another four-quark operator that could contribute is $(\bar{q}^i\gamma_\mu q)(\bar{q}\gamma^\mu q^j)$. However, using the Fierz identity, this can be turned into a linear combination of $O_{qq}^{(1,3)}$ and some other four-quark operators with different isospin and color structures which do not contribute to this process.  Four-quark operators are neglected in the vertex-function approach to the $Wtb$ vertex.

\begin{figure}[tb]
\centering\includegraphics[width=14cm]{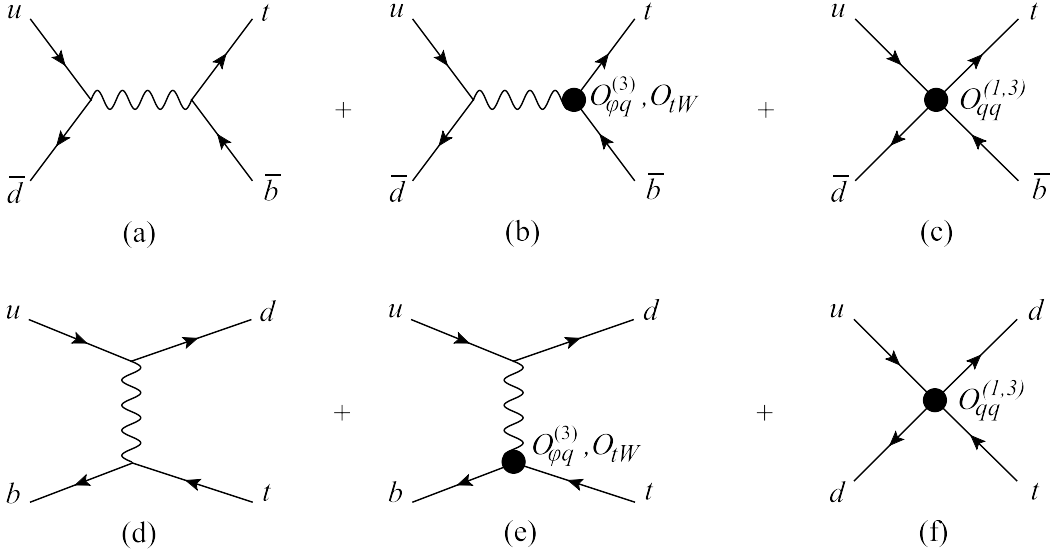}
\caption{Feynman diagrams for the $s$- and $t$-channel single top production. (a-c) are the $s$-channel diagrams, while (d-f) are the $t$-channel diagrams. (a,d) are the SM amplitude, (b,e) are the correction from $O_{\phi q}^{(3)}$ and $O_{tW}$, and (c,f) are the four-fermion interaction from $O_{qq}^{(1,3)}$. The diagrams for the $t$-channel process $\bar{d}b\rightarrow \bar{u}t$ can be obtained by interchanging $u$ and $d$ quarks in (d-f).}\label{fig5}
\end{figure}

The Feynman diagrams are shown in Figure \ref{fig5}.  Since the operator $O_{tW}$ will be measured (or bounded) from studies of top-quark decay, the $s$- and $t$-channel production of single top quarks can be used to measure (or bound) the operators $O_{\phi q}^{(3)}$ and $O_{qq}^{(1,3)}$.

Now we turn to consider the $gb\rightarrow Wt$ process. The contributing operators are
\begin{eqnarray}
O_{\phi q}^{(3)}&=&i(\phi^+\tau^ID_\mu\phi)(\bar{q}\gamma^\mu\tau^Iq)\\
O_{tW}&=&(\bar{q}\sigma^{\mu\nu}\tau^It)\tilde{\phi}W^I_{\mu\nu}\\
O_{tG}&=&(\bar{q}\sigma^{\mu\nu}\lambda^A t)\tilde{\phi}G^A_{\mu\nu}
\end{eqnarray}
Again, the first two operators $O_{\phi q}^{(3)}$ and $O_{tW}$ will affect the $Wtb$ coupling. The ``chromomagnetic moment'' operator
$O_{tG}$ modifies the $gtt$ coupling:
\begin{equation}
L_\mathit{eff}= \frac{{\rm Re}C_{tG}}{{\sqrt 2}\Lambda^2}v\left(\bar{t}\sigma^{\mu\nu}\lambda^A t\right) G_{\mu\nu}^A
\end{equation}
This interaction is neglected in the vertex-function approach to the $Wtb$ vertex.

\begin{figure}[tb]
\centering\includegraphics[width=14cm]{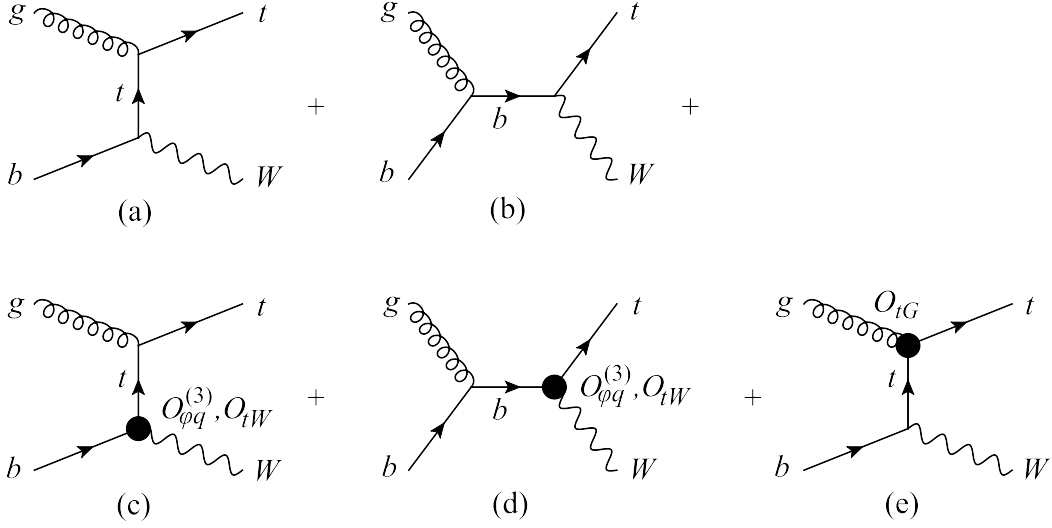}
\caption{The Feynman diagrams for $Wt$ associated production process. (a,b) are the SM amplitude. (c,d) are corrections due to the operator $O_{\phi q}^{(3)}$ and $O_{tW}$. (e) is a modification on the $gtt$ vertex.}\label{fig6}
\end{figure}

The Feynman diagrams are shown in Figure \ref{fig6}.  Since the operators $O_{\phi q}^{(3)}$ and $O_{tW}$ will be measured (or bounded) from single-top production and top-quark decay, respectively, the $Wt$ associated production process can be used to measure (or bound) the operator $O_{tG}$, which is also present in $t\bar t$ production (see Section~\ref{sec:pair}).

\begin{figure}[t]
\begin{minipage}[c]{8cm}
\centering\includegraphics[width=7cm]{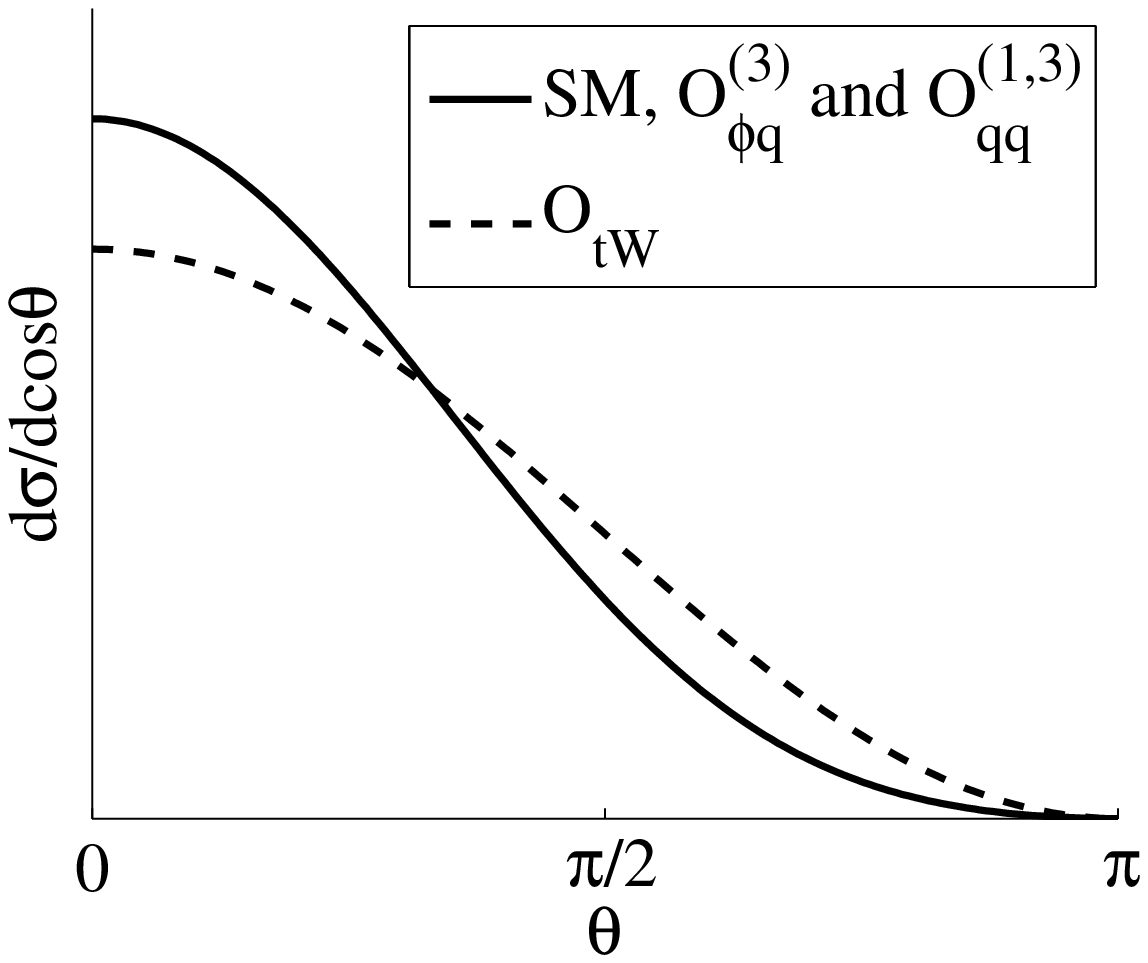}
\caption{The s-channel differential cross section at $\sqrt{s}=2m_t$.}\label{fig7}
\end{minipage}
\hspace{1cm}
\begin{minipage}[c]{8cm}
\centering\includegraphics[width=7cm]{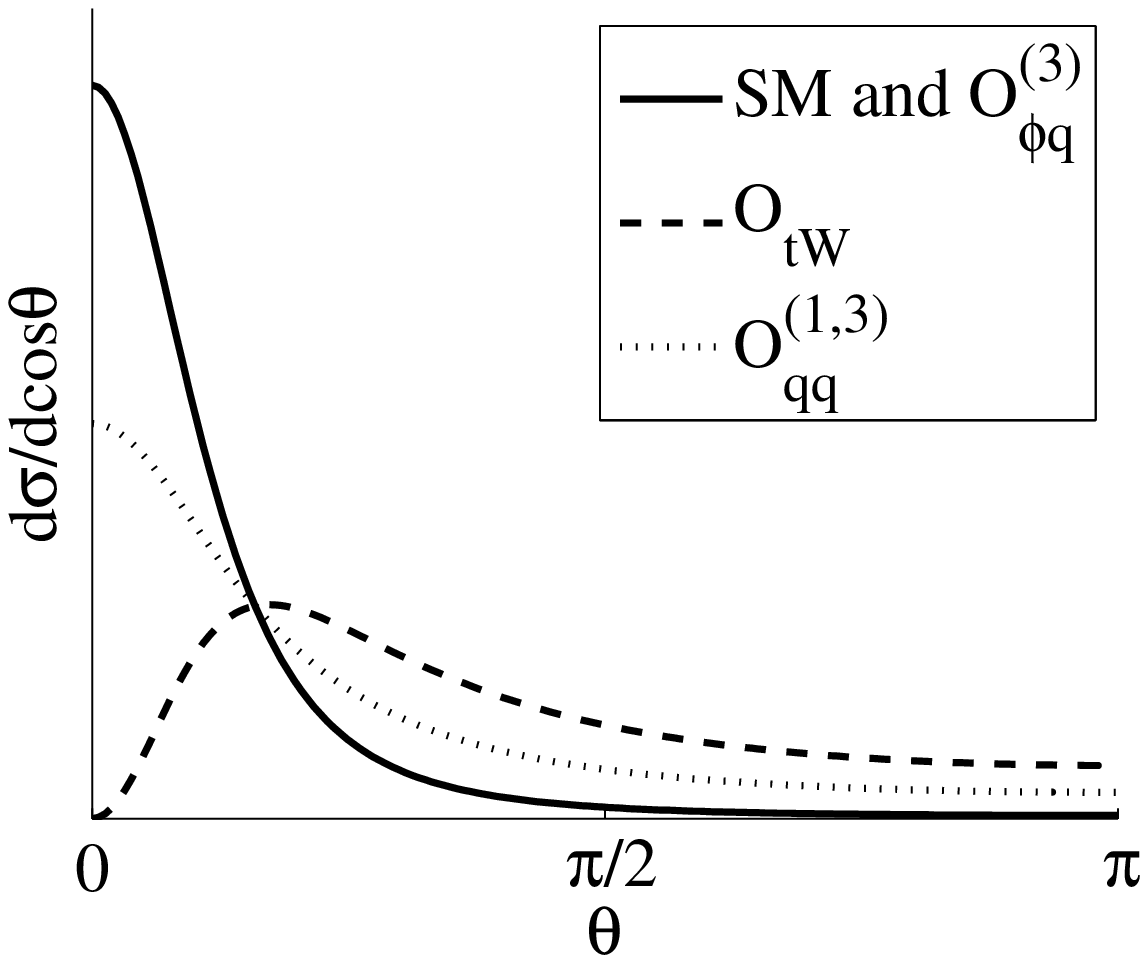}
\caption{The t-channel ($ub\rightarrow dt$) differential cross section at $\sqrt{s}=2m_t$.}\label{fig8}
\end{minipage}
\end{figure}

\begin{figure}[t]
\begin{minipage}[c]{8cm}
\centering\includegraphics[width=7cm]{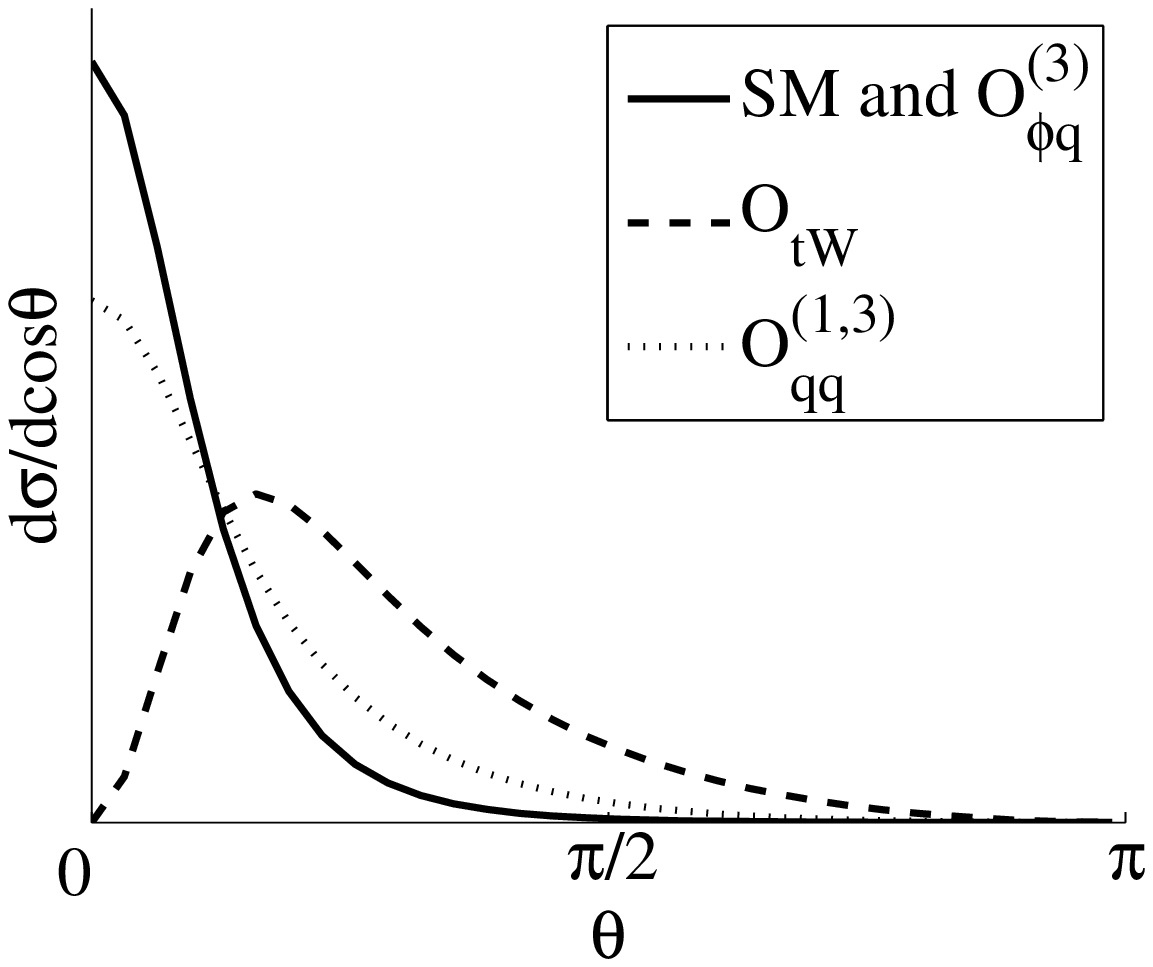}
\caption{The t-channel ($\bar{d}b\rightarrow \bar{u}t$) differential cross section at $\sqrt{s}=2m_t$.}\label{fig9}
\end{minipage}
\hspace{1cm}
\begin{minipage}[c]{8cm}
\centering\includegraphics[width=7cm]{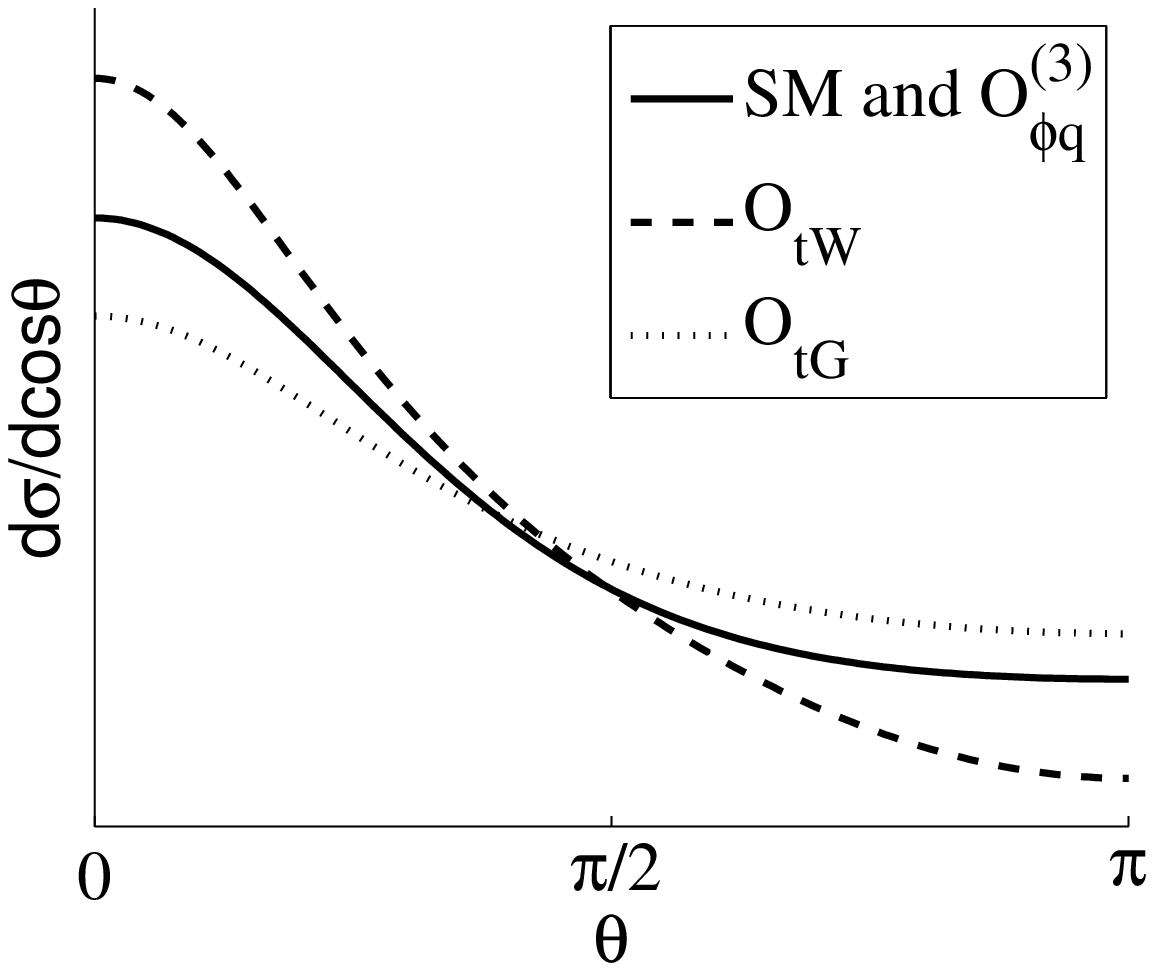}
\caption{The $gb\rightarrow Wt$ channel differential cross section at $\sqrt{s}=2m_t$.}\label{fig10}
\end{minipage}
\end{figure}

Here we list all the corrections to the SM amplitudes and cross sections. The squared amplitude of the three channels are:
\begin{eqnarray}
\mbox{$s$-channel:}\quad&&\nonumber\\
\frac{1}{4}\Sigma|M_{u\bar{d}\rightarrow t\bar{b}}|^2\!\!\!\!\!&=&\!\!\!\!\!\left(V_{tb}^2+\frac{2C_{\phi q}^{(3)}V_{tb}v^2}{\Lambda^2}\right)\frac{g^4u(u-m_t^2)}{4(s-m_W^2)^2}-\frac{2\sqrt{2}{\rm Re}C_{tW}V_{tb}m_tm_W}{\Lambda^2}\frac{g^2su}{(s-m_W^2)^2}+\frac{2C_{qq}^{(1,3)}V_{tb}}{\Lambda^2}\frac{g^2u(u-m_t^2)}{s-m_W^2}\\
\mbox{$t$-channel:}\quad&&\nonumber\\
\frac{1}{4}\Sigma|M_{ub\rightarrow dt}|^2\!\!\!\!\!&=&\!\!\!\!\!\left(V_{tb}^2+\frac{2C_{\phi q}^{(3)}V_{tb}v^2}{\Lambda^2}\right)\frac{g^4s(s-m_t^2)}{4(t-m_W^2)^2}-\frac{2\sqrt{2}{\rm Re}C_{tW}V_{tb}m_tm_W}{\Lambda^2}\frac{g^2st}{(t-m_W^2)^2}+\frac{2C_{qq}^{(1,3)}V_{tb}}{\Lambda^2}\frac{g^2s(s-m_t^2)}{t-m_W^2}\\
\frac{1}{4}\Sigma|M_{\bar{d}b\rightarrow \bar{u}t}|^2\!\!\!\!\!&=&\!\!\!\!\!\left(V_{tb}^2+\frac{2C_{\phi q}^{(3)}V_{tb}v^2}{\Lambda^2}\right)\frac{g^4u(u-m_t^2)}{4(t-m_W^2)^2}-\frac{2\sqrt{2}{\rm Re}C_{tW}V_{tb}m_tm_W}{\Lambda^2}\frac{g^2ut}{(t-m_W^2)^2}+\frac{2C_{qq}^{(1,3)}V_{tb}}{\Lambda^2}\frac{g^2u(u-m_t^2)}{t-m_W^2}
\end{eqnarray}
$Wt$ associated production:
\begin{eqnarray}
\frac{1}{96}\Sigma|M_{gb\rightarrow Wt}|^2&=&\left(V_{tb}^2+\frac{2C_{\phi q}^{(3)}V_{tb}v^2}{\Lambda^2}\right)\frac{g^2g_s^2}{24m_W^2s(t-m_t^2)^2}
\left(m_t^8-(2s+t)m_t^6+((s+t)^2-2tm_W^2-2m_W^4)m_t^4\right.
\nonumber\\&&\left.-(t(s+t)^2-2(s^2-st+2t^2)m_W^2+2tm_W^4-4m_W^6)m_t^2\right.\nonumber\\&&\left.-2tm_W^2(s^2+t^2-2(s+t)m_W^2+2m_W^4)\right)\nonumber\\
&&+\frac{2{\rm Re}C_{tW}V_{tb}g_s^2m_tm_W}{3\sqrt{2}\Lambda^2s(t-m_t^2)^2}
\left(
3m_t^6-(2s+3t+6m_W^2)m_t^4\right.\nonumber\\&&\left.-(s^2+2st-3t^2-6m_W^4)m_t^2+t(s^2-2st-3t^2+6(s+t)m_W^2-6m_W^4)
\right)
\nonumber\\
&&+\frac{{\rm Re}C_{tG}V_{tb}^2g^2g_sm_tv}{3\sqrt{2}\Lambda^2(m_t^2-t)}(m_t^2+2s-t)
\end{eqnarray}
As before, $C_i$ is the coefficient of operator $O_i$ and $s,t,u$ are the usual Mandelstam variables. We have set $V_{ud}=1$ for simplicity.
The differential cross sections are as follows:
\begin{eqnarray}
\frac{d\sigma_{u\bar{d}\rightarrow t\bar{b}}}{d\cos\theta}&=&
\left(V_{tb}^2+\frac{2C_{\phi q}^{(3)}V_{tb}v^2}{\Lambda^2}\right)\frac{g^4(s-m_t^2)^2}{512\pi s^2(s-m_W^2)^2}
(1+\cos\theta)\left(s+m_t^2+(s-m_t^2)\cos\theta\right)\nonumber\\
&&+{\rm Re}C_{tW}V_{tb}\frac{g^2m_tm_W(s-m_t^2)^2}{16\sqrt{2}\pi\Lambda^2s(s-m_W^2)^2}(1+\cos\theta)\nonumber\\
&&+C_{qq}^{(1,3)}V_{tb}\frac{g^2(s-m_t^2)^2}{64\pi\Lambda^2s^2(s-m_W^2)}(1+\cos\theta)(s+m_t^2+(s-m_t^2)\cos\theta)
\label{eq:schannel}\end{eqnarray}
with $\theta$ the angle between up quark and top quark momenta in the center of mass frame;
\begin{eqnarray}
\frac{d\sigma_{ub\rightarrow dt}}{d\cos\theta}\!\!\!&=&\!\!\!
\left(V_{tb}^2+\frac{2C_{\phi q}^{(3)}V_{tb}v^2}{\Lambda^2}\right)\frac{g^4(s-m_t^2)^2}
{32\pi s(2m_W^2+(s-m_t^2)(1-\cos\theta))^2}\nonumber\\
&&\!\!\!\!\!
+{\rm Re}C_{tW}V_{tb}\frac{g^2m_tm_W(s-m_t^2)^2(1-\cos\theta)}{4\sqrt{2}\pi\Lambda^2s(2m_W^2+(s-m_t^2)(1-\cos\theta))^2}-C_{qq}^{(1,3)}V_{tb}\frac{g^2(s-m_t^2)^2}{8\pi\Lambda^2s(2m_W^2+(s-m_t^2)(1-\cos\theta))}
\label{eq:tchannel}\end{eqnarray}

\begin{eqnarray}
\frac{d\sigma_{\bar{d}b\rightarrow \bar{u}t}}{d\cos\theta}&=&
\left(V_{tb}^2+\frac{2C_{\phi q}^{(3)}V_{tb}v^2}{\Lambda^2}\right)
\frac{g^4(s-m_t^2)^2(1+\cos\theta)(s+m_t^2+(s-m_t^2)\cos\theta)}
{128\pi\Lambda^2s^2(2m_W^2+(s-m_t^2)(1-\cos\theta))^2}\nonumber\\
&&
-{\rm Re}C_{tW}V_{tb}\frac{g^2m_tm_W(s-m_t^2)^3\sin^2\theta}
{8\sqrt{2}\pi\Lambda^2 s^2(2m_W^2+(s-m_t^2)(1-\cos\theta))^2}\nonumber\\
&&-C_{qq}^{(1,3)}V_{tb}\frac{g^2(s-m_t^2)^2(1+\cos\theta)(s+m_t^2+(s-m_t^2)\cos\theta)}
{32\pi\Lambda^2 s^2(2m_W^2+(s-m_t^2)(1-\cos\theta))}
\end{eqnarray}
with $\theta$ the angle between bottom quark and top quark momenta in the center of mass frame;
\begin{eqnarray}
\frac{d\sigma_{gb\rightarrow Wt}}{d\cos\theta}\!\!\!\!\!&=&\!\!\!\!\!
\left(V_{tb}^2+\frac{2C_{\phi q}^{(3)}V_{tb}v^2}{\Lambda^2}\right)
\frac{g^2g_s^2\lambda^{1/2}}{1536\pi s^3 m_W^2 (s+m_t^2-m_W^2-\lambda^{1/2}\cos\theta)^2}
\left((m_t^2+10m_W^2)s^3\right.\nonumber\\&&
\left.+(3m_t^4+19m_t^2m_W^2-22m_W^4)s^2-(9m_t^6+8m_t^4m_W^2+5m_t^2m_W^4-22m_W^6)s
+5(m_t^2-m_W^2)^3(m_t^2+2m_W^2)\right.\nonumber\\&&
\left.-(m_t^2+2m_W^2)\lambda^{3/2}\cos^3\theta
+\left((6m_W^2-m_t^2)s-m_t^4-m_t^2m_W^2+2m_W^4\right)\lambda\cos^2\theta
\right.\nonumber\\&&\left.
-\left(
(14m_W^2-m_t^2)s^2-2(m_t^4-7m_t^2m_W^2+6m_W^4)s+3(m_t^6-3m_t^2m_W^4+2m_W^6)
\right)\lambda^{1/2}\cos\theta\right)\nonumber\\&&
-{\rm Re}C_{tW}V_{tb}\frac{g_s^2 m_tm_W \lambda^{1/2}}{96\sqrt{2}\pi\Lambda^2s^3\left(s+m_t^2-m_W^2-\lambda^{1/2}\cos\theta\right)^2}\left(5s^3-9(m_t^2-m_W^2)s^2
\right.\nonumber\\&&
\left.
+(19m_t^4+10m_t^2m_W^2-29m_W^4)s-15(m_t^2-m_W^2)^3+3\lambda^{3/2}\cos^3\theta-(5s-3m_t^2+3m_W^2)\lambda\cos^2\theta
\right.\nonumber\\&&
\left.
-\left(3s^2-10(m_t^2-m_W^2)s-9(m_t^2-m_W^2)^2\right)\lambda^{1/2}\cos\theta
\right)\nonumber\\&&
+{\rm Re}C_{tG}V_{tb}^2\frac{g^2g_svm_t\lambda^{1/2}\left(m_t^2-m_W^2+5s-\lambda^{1/2}\cos\theta\right)}{96\sqrt{2}\pi\Lambda^2s^2
\left(m_t^2-m_W^2+s-\lambda^{1/2}\cos\theta\right)}
\end{eqnarray}
with $\theta$ the angle between gluon and top quark momenta in the center of mass frame, and\\ $\lambda=s^2+m_t^4+m_W^4-2sm_t^2-2sm_W^2-2m_t^2m_W^2$. The angular dependence at $\sqrt{s}=2m_t$ (recall that the kinematic threshold is $\sqrt{s}=m_t$) is shown in Figures \ref{fig7}-\ref{fig10} (areas are normalized).
The total cross sections are:

\begin{eqnarray}
\sigma_{u\bar{d}\rightarrow t\bar{b}}&=&\left(V_{tb}^2+\frac{2C_{\phi q}^{(3)}V_{tb}v^2}{\Lambda^2}\right)
\frac{g^4(s-m_t^2)^2(2s+m_t^2)}{384\pi\Lambda^2s^2(s-m_W^2)^2}
\nonumber\\
&&+{\rm Re}C_{tW}V_{tb}\frac{g^2m_tm_W(s-m_t^2)^2}{8\sqrt{2}\pi\Lambda^2s(s-m_W^2)^2}
+C_{qq}^{(1,3)}V_{tb}\frac{g^2(s-m_t^2)^2(2s+m_t^2)}{48\pi\Lambda^2s^2(s-m_W^2)}
\end{eqnarray}

\begin{eqnarray}
\sigma_{ub\rightarrow dt}&=&\left(V_{tb}^2+\frac{2C_{\phi q}^{(3)}V_{tb}v^2}{\Lambda^2}\right)
\frac{g^4(s-m_t^2)^2}{64\pi\Lambda^2sm_W^2(s-m_t^2+m_W^2)}
\nonumber\\
&&-{\rm Re}C_{tW}V_{tb}\frac{g^2m_tm_W\left(s-m_t^2-(s-m_t^2+m_W^2)\ln\frac{s-m_t^2+m_W^2}{m_W^2}\right)}{4\sqrt{2}\pi\Lambda^2s(s-m_t^2+m_W^2)}-C_{qq}^{(1,3)}V_{tb}\frac{g^2(s-m_t^2)\ln\frac{s-m_t^2+m_W^2}{m_W^2}}{8\pi\Lambda^2s}
\end{eqnarray}

\begin{eqnarray}
\sigma_{\bar{d}b\rightarrow \bar{u}t}&=&\left(V_{tb}^2+\frac{2C_{\phi q}^{(3)}V_{tb}v^2}{\Lambda^2}\right)
\frac{g^4\left((s+2m_W^2)(s-m_t^2)-m_W^2(2s+2m_W^2-m_t^2)\ln\frac{s+m_W^2-m_t^2}{m_W^2}\right)}{64\pi s^2m_W^2}\nonumber\\
&&
-{\rm Re}C_{tW}V_{tb}\frac{g^2m_tm_W\left((s+2m_W^2-m_t^2)\ln\frac{s+m_W^2-m_t^2}{m_W^2}-2(s-m_t^2)\right)}{4\sqrt{2}\pi\Lambda^2 s^2}\nonumber\\&&
-C_{qq}^{(1,3)}V_{tb}\frac{g^2\left(2(s+m_W^2-m_t^2)(s+m_W^2)\ln\frac{s+m_W^2-m_t^2}{m_W^2}
-(s-m_t^2)(3s+2m_W^2-m_t^2)\right)}
{16\pi\Lambda^2 s^2}
\end{eqnarray}

\begin{eqnarray}
\sigma_{gb\rightarrow Wt}&=&
\left(V_{tb}^2+\frac{2C_{\phi q}^{(3)}V_{tb}v^2}{\Lambda^2}\right)
\frac{g^2g_s^2}{384s^3m_W^2}
\biggl(
-\left((3m_t^2-2m_W^2)s+7(m_t^2-m_W^2)(m_t^2+2m_W^2)\right)\lambda^{1/2}\biggr.
\nonumber\\&&\left.+2(m_t^2+2m_W^2)(s^2+2(m_t^2-m_W^2)s+2(m_t^2-m_W^2)^2)\ln
\left(\frac{s+m_t^2-m_W^2+\lambda^{1/2}}{s+m_t^2-m_W^2-\lambda^{1/2}}\right)
\right)\nonumber\\&&
-{\rm Re}C_{tW}V_{tb}\frac{g_s^2m_tm_W}{24\sqrt{2}\Lambda^2s^3}
\biggl(\left(s+21(m_t^2-m_W^2)\right)\lambda^{1/2}\biggr.\nonumber\\&&
\left.
+2\left(s^2-6(m_t^2-m_W^2)s-6(m_t^2-m_W^2)^2\right)
\ln
\left(\frac{s+m_t^2-m_W^2+\lambda^{1/2}}{s+m_t^2-m_W^2-\lambda^{1/2}}\right)\right)\nonumber\\&&
+{\rm Re}C_{tG}V_{tb}^2\frac{g^2g_s vm_t}{24\sqrt{2}\Lambda^2s^2}
\left(
2s\ln
\left(\frac{s+m_t^2-m_W^2+\lambda^{1/2}}{s+m_t^2-m_W^2-\lambda^{1/2}}\right)
+\lambda^{1/2}
\right)
\end{eqnarray}

The operators $O_{\phi q}^{(3)}$ and $O_{qq}^{(1,3)}$ will be measured (or bounded) by single top production.  Because they enter with the opposite relative sign in $s$- and $t-$ channel production (see Eqs.~(\ref{eq:schannel}),(\ref{eq:tchannel})), it will be valuable to measure these two processes separately.

The operator $O_{tW}$ also has an effect on the produced top quark spin.
In the SM $s-$ and $t-$channel single top production,
the top quark is always polarized in the direction of $d$ or $\bar{d}$
three-momentum in the top rest frame \cite{Mahlon:1996pn}.
When $O_{tW}$ is present, the top quark spin devitates from its original direction,
but is still in the production plane.
For example, in the $s-$channel process, the top spin deviates away from the
three-momentum of the $\bar{b}$, with an angle (in the top rest frame)
\begin{equation}\label{eq:topspin1}
\delta\theta={\rm Re}C_{tW}\frac{2\sqrt{2}v^2}{\Lambda^2}\frac{\sqrt{s}}{m_W}
\frac{(s-m_t^2)\sin\theta}{s+m_t^2+(s-m_t^2)\cos\theta}
\end{equation}
where $\theta$ is the scattering angle in the $W$ rest frame.
Similarly, in the $t-$channel process $b\bar{d}\rightarrow t\bar{u}$,
the top spin deviates toward the three-momentum of $\bar{u}$, with the same angle.
In the $t-$channel process $bu\rightarrow td$, the top spin deviates toward
the three-momentum of the incoming $b$ quark, with an angle
\begin{equation}\label{eq:topspin2}
\delta\theta={\rm Re}C_{tW}\frac{\sqrt{2}v^2}{\Lambda^2}\frac{\sqrt{s}}{m_W}\sin\theta
\end{equation}

In Eq.~(\ref{eq:anapower1}) we reported the effect of the operator $O_{tW}$ on the analyzing power of top decay.
Let $\hat{\mathbf s}$ be the unit vector in the top quark spin direction
and $\hat{\mathbf p}_i$ be the unit vector in the direction of the three-momentum
of the decay product $i$ in the top rest frame, we have
\begin{equation}
\alpha_i=3<\hat{\mathbf s}\cdot \hat{\mathbf p}_i>
\end{equation}
In practice, we can use the $s-$ and $t-$channel single top production
as a source of polarized top quark. To measure the analyzing power,
we can replace $\hat{\mathbf s}$
with $\hat{\mathbf p}_{d,\bar{d}}$, the unit vector in the direction of three-momentum
of $d$ or $\bar{d}$, depending on the production channel:
\begin{equation}\label{eq:anapower2}
\alpha_i=3<\hat{\mathbf p}_{d,\bar{d}}\cdot \hat{\mathbf p}_i>
\end{equation}
In single top production the top quark spin is affected by $O_{tW}$, so
$\hat{\mathbf s}$ and $\hat{\mathbf p}_{d,\bar{d}}$ are not exactly aligned.
However, the direction in which the top quark spin deviates from the three-momentum of
$d$ or $\bar{d}$ is independent of the $\hat{\mathbf p}_i$, i.e.
\begin{equation}\label{eq:averageout}
<(\hat{\mathbf p}_{d,\bar{d}}-\hat{\mathbf s})\cdot \hat{\mathbf p}_i>=0
\end{equation}
Therefore Eq.~(\ref{eq:anapower2}) still holds. In other words, the effect of $O_{tW}$
on the production vertex doesn't affect the measurement of the analyzing power.

\section{Top Pair Production}\label{sec:pair}

The effect of higher dimension operators on top quark pair production is studied in \cite{Cho:1994yu,Kumar:2009vs,Lillie:2007hd}. In Ref.~\cite{Cho:1994yu}, two dimension-six operators, the chromomagnetic moment operator, $O_{tG}$, and the triple gluon field strength operator, $O_{G}$, are considered:
\begin{eqnarray}
O_{tG}&=&(\bar{q}\sigma^{\mu\nu}\lambda^A t)\tilde{\phi}G^A_{\mu\nu}\\
O_{G}&=&f_{ABC}G^{A\nu}_\mu G^{B\rho}_\nu G^{C\mu}_\rho
\end{eqnarray}
It is shown that $O_{G}$ will generate observable cross section deviations from QCD at the LHC even for relatively small values of its coefficient.

Here we redo the leading order calculation, and also take into account the operator $O_{\phi G}$:
\begin{eqnarray}
O_{\phi G}=\frac{1}{2}(\phi^+\phi)G^A_{\mu\nu}G^{A\mu\nu}
\end{eqnarray}
which is a Higgs-gluon interaction. Its effect on the Higgs production rate and branching ratios has been discussed in \cite{Manohar:2006gz}. We include this operator because it contributes to top pair production through $gg\rightarrow h\rightarrow t\bar{t}$,
\begin{equation}
L_\mathit{eff}= \frac{1}{2}\frac{C_{\phi G}}{\Lambda^2}vhG^A_{\mu\nu}G^{A\mu\nu}
\end{equation}
which could be significant because the top quark has a large Yukawa coupling.

\begin{figure}[tb]
\centering\includegraphics[width=12.5cm]{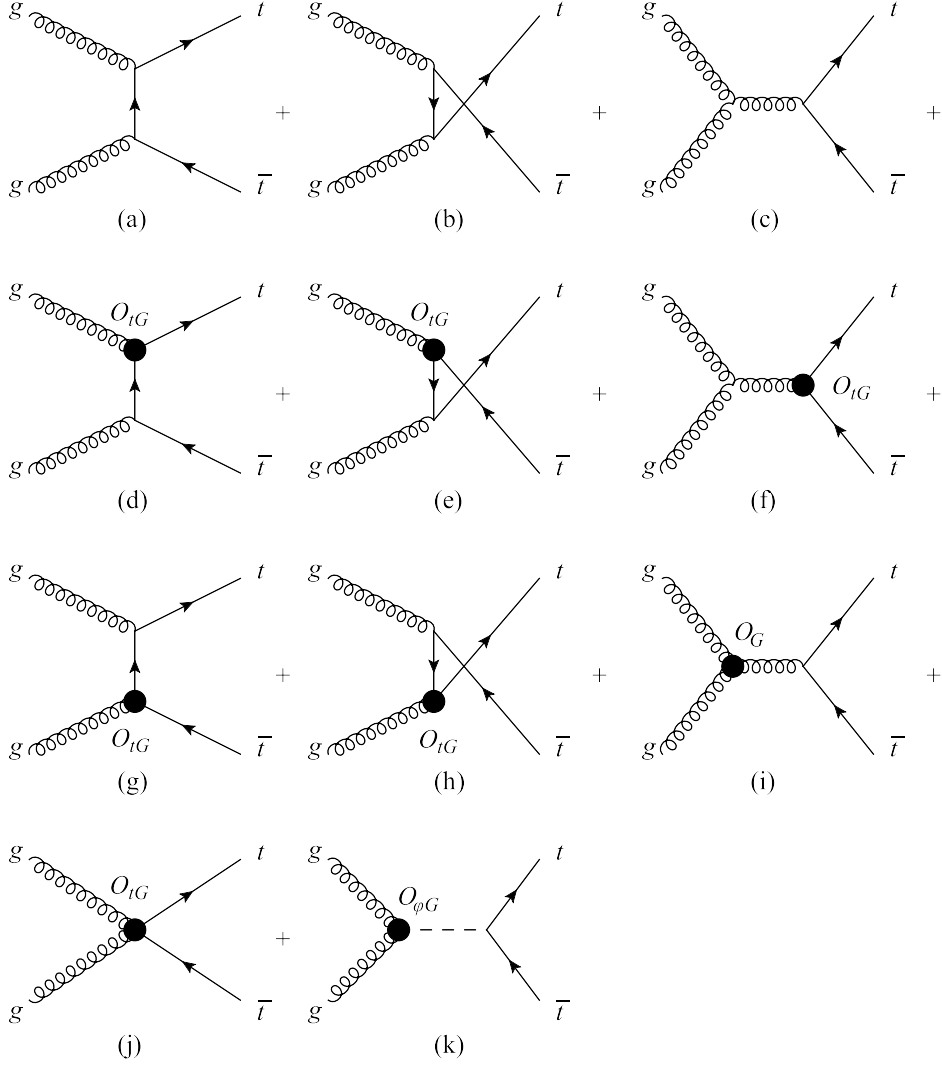}
\caption{The Feynman diagrams for $gg\rightarrow t\bar{t}$ process. Diagram (a-c) are the SM amplitude. (d-h) are the $gtt$ vertex correction induced by $O_{tG}$. (i) is the $g^3$ vertex correction induced by $O_{G}$. (j) is a $ggtt$ interaction from $O_{tG}$, and (k) is a $gg\rightarrow h\rightarrow tt$ process, induced by $O_{\phi G}$.}\label{fig11}
\end{figure}

Top quark pair production proceeds at the tree level through the parton reactions
$gg\rightarrow t\bar{t}$ and $q\bar{q}\rightarrow t\bar{t}$. We first consider the gluon channel. The Feynman diagrams are shown in Figure \ref{fig11}. The operator $O_{tG}$ changes the SM $gtt$ coupling, and also generates a new $ggtt$ interaction. $O_{G}$ affects the three-point gluon vertex in QCD. $O_{\phi G}$ generates a new diagram with an $s$-channel Higgs boson.

The squared amplitude is:
\begin{eqnarray}
\frac{1}{256}|M|^2&=&\frac{3g_s^4}{4}\frac{(m^2-t)(m^2-u)}{s^2}-\frac{g_s^4}{24}\frac{m^2(s-4m^2)}{(m^2-t)(m^2-u)}+\frac{g_s^4}{6}\frac{tu-m^2(3t+u)-m^4}{(m^2-t)^2}\nonumber\\&&
+\frac{g_s^4}{6}\frac{tu-m^2(t+3u)-m^4}{(m^2-u)^2}
-\frac{3g_s^4}{8}\frac{tu-2m^2t+m^4}{s(m^2-t)}-\frac{3g_s^4}{8}\frac{tu-2m^2u+m^4}{s(m^2-u)}
\nonumber\\&&
+\frac{\sqrt{2}{\rm Re}C_{tG}g_s^3vm}{3\Lambda^2}\frac{4s^2-9tu-9m^2s+9m^4}{(m^2-t)(m^2-u)}
+\frac{9C_{G}g_s^3}{8\Lambda^2}\frac{m^2(t-u)^2}{(m^2-t)(m^2-u)}
\nonumber\\&&
-\frac{C_{\phi G}g_s^2m^2}{8\Lambda^2}\frac{s^2(s-4m^2)}{(s-m_h^2)(t-m^2)(u-m^2)}
\end{eqnarray}
where $m$ is the mass of the top quark and $m_h$ is the mass of the Higgs boson.

The differential and total cross sections are
\begin{eqnarray}
\frac{d\sigma}{d\cos\theta}
\!\!\!\!\!&=&\!\!\!\!\!
\frac{g_s^4\beta}{1536\pi s(1-\beta^2\cos^2\theta)^2}\nonumber\\
&&\!\!\!\!\!
\left(
7(1+2\beta^2-2\beta^4)-\beta^2(5-32\beta^2+18\beta^4)\cos^2\theta-(25\beta^4-18\beta^6)\cos^4\theta-9\beta^6\cos^6\theta
\right)\nonumber\\
&&\!\!\!\!\!+{\rm Re}C_{tG}\frac{g_s^3v\beta\sqrt{1-\beta^2}(7+9\beta^2\cos^2\theta)}{96\sqrt{2}\pi\Lambda^2 \sqrt{s}(1-\beta^2\cos^2\theta)}
+C_{G}\frac{9g_s^3\beta^3(1-\beta^2)\cos^2\theta}{256\pi\Lambda^2(1-\beta^2\cos^2\theta)}
-C_{\phi G}\frac{g_s^2s\beta^3(1-\beta^2)}{256\pi\Lambda^2(s-m_h^2)(1-\beta^2\cos^2\theta)}
\end{eqnarray}

\begin{eqnarray}
\sigma&=&\frac{g_s^4}{768\pi s}\left(31\beta^3-59\beta+(33-18\beta^2+\beta^4)\ln\frac{1+\beta}{1-\beta}\right)
+{\rm Re}C_{tG}\frac{g_s^3v\sqrt{1-\beta^2}}{48\sqrt{2}\pi\Lambda^2\sqrt{s}}
\left(8\ln\frac{1+\beta}{1-\beta}-9\beta\right)\nonumber\\
&&+C_{G}\frac{9g_s^3(1-\beta^2)}{256\pi\Lambda^2}\left(\ln\frac{1+\beta}{1-\beta}-2\beta\right)
-C_{\phi G}\frac{g_s^2s\beta^2(1-\beta^2)}{256\pi\Lambda^2(s-m_h^2)}\ln\frac{1+\beta}{1-\beta}
\end{eqnarray}
Here $\theta$ is the angle between the gluon and top quark momenta in the center of mass frame; $\beta\equiv \sqrt{1-\frac{4m^2}{s}}$ is the velocity of the top quark. Top quark pair production can be used to measure (or bound) the coefficients of the operators $O_{tG}$, $O_{G}$ and $O_{\phi G}$.  The operator $O_{tG}$ is also probed by $Wt$ associated production, as discussed above, and the operator $O_{\phi G}$ is probed by Higgs production \cite{Manohar:2006gz}.

Now we turn to consider the quark process $q\bar{q}\rightarrow t\bar{t}$. There are a large number of four-quark operators with different chiral and flavor structures \cite{Leung:1984ni,Buchmuller:1985jz,Cho:1994yu}. Here we consider all possible chirality and color structures. In Ref.~\cite{Buchmuller:1985jz}, only one generation is considered. When there are three generations, the quark field in these operators can be of any generation. For example, $(\bar{q}^i\gamma_\mu q^j)(\bar{q}\gamma^\mu q)$
and
$(\bar{q}^i\gamma_\mu q)(\bar{q}\gamma^\mu q^j)$
(superscripts $i,j$ are used to denote the first two generations)
should be considered as different operators. The effect of some of these operators are suppressed by the color structure or by the small quark mass. For example, $(\bar{q}^i\gamma_\mu q^j)(\bar{q}\gamma^\mu q)$ doesn't interfere with the SM, because the $t$ and $\bar{t}$ form a color singlet; an operator like $(\bar{q}t)\epsilon(\bar{q}^id^j)$ doesn't interfere either, because it involves a left-handed and a right-handed down quark while the SM $gd\bar{d}$ coupling doesn't change chirality.

\begin{figure}[tb]
\centering\includegraphics[width=13.5cm]{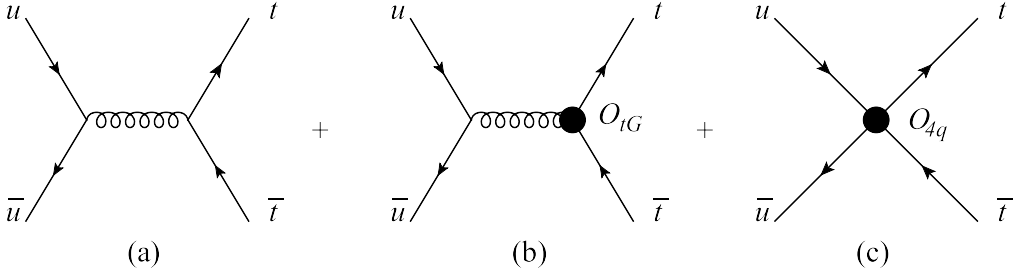}
\caption{The Feynman diagrams for $u\bar{u}\rightarrow t\bar{t}$ process. (a) is the SM amplitude, (b) is the correction on $gtt$ coupling induced by $O_{tG}$, and (c) is the four-fermion interactions. The $d\bar{d}\rightarrow t\bar{t}$ process has the same diagrams.}\label{fig12}
\end{figure}

Using the Fierz transformation and the following SU(2) and SU(3) identities
\begin{eqnarray}
\tau^I_{ab}\tau^I_{cd}&=&-\delta_{ab}\delta_{cd}+2\delta_{ad}\delta_{bc}\nonumber\\
t^A_{ij}t^A_{kl}&=&-\frac{1}{6}\delta_{ij}\delta_{kl}+\frac{1}{2}\delta_{il}\delta_{jk}
\end{eqnarray}
we find that only the following four-quark operators contribute to the $u\bar{u},d\bar{d}\rightarrow t\bar{t}$ reaction:
\begin{equation}
\begin{array}{ll}
O^{(8,1)}_{qq}=\frac{1}{4}(\bar{q}^i\gamma_{\mu} \lambda^A q^j)(\bar{q}\gamma^{\mu} \lambda^A q) &
O^{(8,3)}_{qq}=\frac{1}{4}(\bar{q}^i\gamma_{\mu} \tau^I \lambda^A q^j)(\bar{q}\gamma^{\mu} \tau^I \lambda^A q)\\
O^{(8)}_{ut}=\frac{1}{4}(\bar{u}^i\gamma_\mu \lambda^A u^j)(\bar{t} \gamma^\mu \lambda^A t) &
O^{(8)}_{dt}=\frac{1}{4}(\bar{d}^i\gamma_\mu \lambda^A d^j)(\bar{t} \gamma^\mu \lambda^A t)\\
O^{(1)}_{qu}=(\bar{q}u^i)(\bar{u}^jq) &
O^{(1)}_{qd}=(\bar{q}d^i)(\bar{d}^jq)\\
O^{(1)}_{qt}=(\bar{q}^it)(\bar{t}q^j)
\end{array}
\end{equation}
We don't include the operators that have the form $(\bar{q}\lambda^A u^i)(\bar{u}^j\lambda^Aq)$. This operator can be turned into a linear combination of $O^{(1)}_{qu}$, which is already considered, and another operator $(\bar{q}_cu^i_b)(\bar{u}^j_aq_d)\delta_{ab}\delta_{cd}$ ($a,b,c,d$ denote color indices), which does not contribute because the $t$ and $\bar{t}$ form a color singlet. In addition, we also need to consider the operator $O_{tG}$, whose effect is to change the $gtt$ coupling. The diagrams are shown in Figure \ref{fig12}.
The result is
\begin{eqnarray}
\frac{1}{36}|M_{\bar{u}u\rightarrow \bar{t}t}|^2&=&g_s^2(M_1^2+M_2^2)+\frac{32\sqrt{2}{\rm Re}C_{tG}g_s^3vm}{9\Lambda^2}
+C_u^1\frac{s}{\Lambda^2}M_1^2
+C_u^2\frac{s}{\Lambda^2}M_2^2\nonumber\\
\frac{1}{36}|M_{\bar{d}d\rightarrow \bar{t}t}|^2&=&g_s^2(M_1^2+M_2^2)+\frac{32\sqrt{2}{\rm Re}C_{tG}g_s^3vm}{9\Lambda^2}
+C_d^1\frac{s}{\Lambda^2}M_1^2
+C_d^2\frac{s}{\Lambda^2}M_2^2
\end{eqnarray}
where
\begin{eqnarray}
C_u^1&=&C^{(8,1)}_{qq}+C^{(8,3)}_{qq}+C^{(8)}_{ut}\\
C_u^2&=&C^{(1)}_{qu}+C^{(1)}_{qt}\\
C_d^1&=&C^{(8,1)}_{qq}-C^{(8,3)}_{qq}+C^{(8)}_{dt}\\
C_d^2&=&C^{(1)}_{qd}+C^{(1)}_{qt}\\
M_1^2&=&\frac{4g_s^2}{9s^2}(3m^4-m^2(t+3u)+u^2)\\
M_2^2&=&\frac{4g_s^2}{9s^2}(3m^4-m^2(3t+u)+t^2)
\end{eqnarray}
The cross section is
\begin{eqnarray}
\frac{d\sigma_{\bar{u}u,\bar{d}d\rightarrow \bar{t}t}}{d\cos\theta}&=&
\frac{g_s^4}{144\pi s}\beta(2-\beta^2\sin^2\theta)
+{\rm Re}C_{tG}\frac{g_s^3v\beta\sqrt{1-\beta^2}}{9\sqrt{2}\pi\Lambda^2\sqrt{s}}\nonumber\\&&
+C_{u,d}^1\frac{g_s^2}{288\pi\Lambda^2}\beta(2+2\beta\cos\theta-\beta^2\sin^2\theta)
\nonumber\\&&
+C_{u,d}^2\frac{g_s^2}{288\pi\Lambda^2}\beta(2-2\beta\cos\theta-\beta^2\sin^2\theta)
\end{eqnarray}
where $\theta$ is the angle between up or down quark and the top quark momenta, in the center of mass frame. The total cross section is
\begin{eqnarray}
\sigma_{\bar{u}u,\bar{d}d\rightarrow \bar{t}t}&=&\frac{g_s^4}{108\pi s}\beta(3-\beta^2)
+{\rm Re}C_{tG}\frac{\sqrt{2}g_s^3v}{9\pi\Lambda^2\sqrt{s}}\beta\sqrt{1-\beta^2}
+(C_{u,d}^1+C_{u,d}^2)\frac{g_s^2}{216\pi\Lambda^2}\beta(3-\beta^2)
\end{eqnarray}
Although there are seven four-fermion operators, their effects on top-quark pair production are summarized by only four coefficients $C_{u,d}^{1,2}$.  Thus top-quark pair production can be used to bound four linear combinations of the four-quark operators as well as the operator $O_{tG}$.

If $C_{u,d}^1$ and $C_{u,d}^2$ are distinct, they will generate a forward-backward asymmetry:
\begin{eqnarray}
A^t_{FB}&=&\frac{N(\cos\theta>0)-N(\cos\theta<0)}{N(\cos\theta>0)+N(\cos\theta<0)}\nonumber\\
&=&(C^1_{u,d}-C^2_{u,d})\frac{3s\beta}{4g_s^2\Lambda^2(3-\beta^2)}
\end{eqnarray}
The recent measurements of the top quark forward-backward asymmetry from the CDF and the D0
experiments can be found in \cite{Aaltonen:2008hc,Stricker,CDF,:2007qb,D0}. The SM prediction
is dominated by
$O(\alpha_S^3)$ QCD interference effects and is $5\%$ in the lab frame \cite{Kuhn:1998jr,Kuhn:1998kw,Bowen:2005ap,Almeida:2008ug}. There is a discrepancy of about
$2\sigma$ between theory and experiment. It is interesting to ask whether this discrepancy can be accommodated within the effective field theory framework.
The challenge is to avoid too large a modification of the $t\bar{t}$ production cross section, since
the current measurement is in good agreement with the SM prediction \cite{Aaltonen}. In the effective field theory approach, this can be done if $C_{u,d}^1$ and $C_{u,d}^2$ have similar non-zero values but with opposite sign, i.e.~$C_{u,d}^1\approx-C_{u,d}^2$.

\section{CP Violation}

Violations of the CP symmetry are of great interest in particle
physics especially since its origin is still unclear. Better understanding of this rare phenomenon can lead to
new physics which may explain both the origin of mass and the preponderance of matter over anti-matter
in the present universe.

The SM predicts that CP-violating effects in top physics are very small. This is primarily due
to the fact that its large mass renders the Glashow-Iliopoulos-Maiani (GIM) \cite{Glashow:1970gm} cancellation particularly
effective \cite{Eilam:1990zc,Grzadkowski:1990sm}. Therefore, the study of CP-violation effects in top physics is important because any observation of such effects would be a clear evidence of physics beyond the SM.

Effective field theory is a complete and model-independent approach to physics beyond the SM. Its CP-odd operators can be used to describe the CP-violation effects in top quark physics. We find that there are four CP-odd operators that can have significant contribution to top quark production and decay processes, as listed in Table 2. In this section we will consider the effects of these four operators.

\subsection{Polarized Top Quark Decay}

In top quark decay, the momenta of the four particles, $t$,$b$,$e^+$ and $\nu$ are not independent because of the energy-momentum conservation. However, if we define the top quark spin vector (in the top rest frame):
\begin{equation}
s=(0,\hat{\mathbf{s}})
\end{equation}
where the unit vector $\hat{\mathbf{s}}$ is the direction of the top quark spin, then a term proportional to $\epsilon_{\mu\nu\rho\sigma}p_t^\mu p_b^\nu p_{e^+}^\rho s^\sigma$ is $T_N$-odd. Thus it becomes possible to observe CP violation effects.

In the top quark decay process, there is only one operator that contributes at leading order:
\begin{equation}
O_{tW}=(\bar{q}\sigma^{\mu\nu}\tau^It)\tilde{\phi}W^I_{\mu\nu}
\end{equation}
This operator is CP-odd if its coefficient is imaginary.

To investigate the effect of $O_{tW}$, we choose the coordinate axes in the top rest frame such that the positron momentum is in the $z$-direction, and the bottom momentum is in the $xz$ plane, with a positive $x$ component. The top quark spin is $\hat{\mathbf{s}}=(\sin\theta\cos\phi,\sin\theta\sin\phi,\cos\theta)$.
The decay rate is given by:
\begin{equation}
\frac{d\Gamma}{d\cos\theta d\phi}=
\frac{V_{tb}^2g^4(m_t^6-3m_W^4m_t^2+2m_W^6)}{12288\pi^3m_t^3m_W\Gamma_W}(1+\cos\theta)
-
\frac{{\rm Im}C_{tW}V_{tb}g^2m_W(m_t^2-m_W^2)^3}
{2048\sqrt{2}\pi^2\Lambda^2\Gamma_W m_t^3}
\sin\theta\sin\phi
\end{equation}
The CP-odd contribution is proportional to $\sin\phi$, so it doesn't affect the total decay rate and the analyzing power $\alpha_i$ defined in Eq.~(\ref{define_alpha}).

We now define the following triple-product and evaluate it in the top rest frame:
\begin{equation}\label{T1}
T=-\frac{1}{m_t}\epsilon_{\mu\nu\rho\sigma}p_t^\mu p_b^\nu p_{e^+}^\rho s^\sigma=(\mathbf{p}_b\times\mathbf{p}_{e^+})\cdot \hat{\mathbf{s}}
\end{equation}
which corresponds to the projection of the top spin onto the direction perpendicular to the plane formed by the bottom quark and the positron.
This leads to an asymmetry:
\begin{eqnarray}
A_{t\to Wb}&=&\frac{N(T>0)-N(T<0)}{N(T>0)+N(T<0)}\nonumber\\
&=&{\rm Im}C_{tW}\frac{3\pi v^2(m_t^2-m_W^2)}{4\sqrt{2}\Lambda^2V_{tb}(m_t^2+2m_W^2)}
\label{eq:tdecayasymm}\end{eqnarray}
Such an asymmetry is a sign of CP violation.  To observe such an asymmetry requires a source of polarized top quarks.  This is addressed in the next section.

\subsection{Spin Asymmetry in Single Top Production}

In single top production, we can construct CP-odd observables in a similar way. In the $s$- and $t$-channel processes, $O_{tW}$ (with imaginary coefficient) is the only CP-odd operator that contributes. Consider the $s$-channel process $u\bar{d}\rightarrow t\bar{b}$. We can define the following triple-product in the top rest frame
\begin{equation}
T=-\frac{1}{m_t}
\epsilon_{\mu\nu\rho\sigma}p_t^\mu p_u^\nu p_{\bar{d}}^{\rho} s^\sigma
=(\mathbf{p}_u\times\mathbf{p}_{\bar{d}})\cdot\mathbf{\hat{s}}
\end{equation}

In the SM, the top spin in its rest frame is in the direction of the $\bar{d}$ three-momentum \cite{Mahlon:1996pn}, therefore $T=0$. When the CP-odd operator is added, the direction of the top quark spin can be computed. It deviates from the production plane, with an angle (in the top rest frame)
\begin{equation}
\theta={\rm Im}C_{tW}\frac{2\sqrt{2}v^2\sqrt{s}(s-m_t^2)\sin\theta_W}
{\Lambda^2V_{tb} m_W(s+m_t^2+(s-m_t^2)\cos\theta_W)}
\end{equation}
where $\theta_W$ is the angle between the momenta of the up quark and the top quark in the $W$ rest frame. The value of $T$ is then given by
\begin{equation}
T=-\frac{\sqrt{2}{\rm Im}C_{tW}v^2s(s-m_t^2)^2\sin^2\theta_W}
{2\Lambda^2V_{tb}m_Wm_t[s+m_t^2+(s-m_t^2)\cos\theta_W]}
\end{equation}

In practice, assume the top spin $\mathbf{\hat{s}}$ is measured in the direction perpendicular to the production plane,
i.e.~$\mathbf{s}_\perp$ takes either $1$ or $-1$, then this will lead to an asymmetry:
\begin{eqnarray}
A_{u\bar{d}\rightarrow t\bar{b}}&=&\frac{N(\mathbf{s}_\perp=1)-N(\mathbf{s}_\perp=-1)}{N(\mathbf{s}_\perp=1)+N(\mathbf{s}_\perp=-1)}\nonumber\\
&=&{\rm Im}C_{tW}
\frac{3\pi v^2\sqrt{s}(s-m_t^2)}{2\sqrt{2}\Lambda^2V_{tb}m_W(2s+m_t^2)}
\label{eq:schannelasymm}\end{eqnarray}
Similarly, for the $t$-channel process $bu\rightarrow td$, we find
\begin{equation}
T=
(\mathbf{p}_b\times\mathbf{p}_{u})\cdot\hat{\mathbf{s}}
=\frac{{\rm Im}C_{tW}v^2s(s-m_t^2)\sin^2\theta_W}{2\sqrt{2}\Lambda^2V_{tb} m_W m_t}
\end{equation}
and for the process $b\bar{d} \rightarrow t\bar{u}$,
\begin{equation}
T=
(\mathbf{p}_b\times\mathbf{p}_{\bar{d}})\cdot\hat{\mathbf{s}}
=\frac{\sqrt{2}{\rm Im}C_{tW}v^2s(s-m_t^2)^2\sin^2\theta_W}{2\Lambda^2V_{tb}m_Wm_t(s+m_t^2+(s-m_t^2)\cos\theta_W)}
\end{equation}
If the top spin $\mathbf{\hat{s}}$ is measured in the direction perpendicular to the production plane,
the corresponding asymmetries are
\begin{eqnarray}
A_{bu\rightarrow td}
&=&\frac{N(\mathbf{s}_\perp=1)-N(\mathbf{s}_\perp=-1)}{N(\mathbf{s}_\perp=1)+N(\mathbf{s}_\perp=-1)}\nonumber\\
&=&-{\rm Im}C_{tW}
\frac{\sqrt{2}\pi v^2\sqrt{s}((s-m_t^2+2m_W^2)\sqrt{s-m_t^2+m_W^2}-2m_W(s-m_t^2+m_W^2))}
{\Lambda^2V_{tb}(s-m_t^2)^2}\label{eq:t1channelasymm}
\\
A_{b\bar{d} \rightarrow t\bar{u}}
&=&\frac{N(\mathbf{s}_\perp=1)-N(\mathbf{s}_\perp=-1)}{N(\mathbf{s}_\perp=1)+N(\mathbf{s}_\perp=-1)}\nonumber\\
&=&-{\rm Im}C_{tW}
\frac{\sqrt{2}\pi v^2\sqrt{s}((s-m_t^2+4m_W^2)\sqrt{s-m_t^2+mW^2}-(3s-3m_t^2+4m_W^2)m_W)}{\Lambda^2V_{tb}((s-m_t^2)(s+2m_W^2)-m_W^2(2s+2m_W^2-m_t^2)\ln\frac{s-m_t^2+m_W^2}{m_W^2})}
\label{eq:t2channelasymm}\end{eqnarray}

In $Wt$ associated production channel $gb\rightarrow Wt$, the chromo-electric dipole moment operator
\begin{equation}
O_{tG}=(\bar{q}\sigma^{\mu\nu}\lambda^A t)\tilde{\phi}G^A_{\mu\nu}
\end{equation}
will also contribute. We find
\begin{eqnarray}\label{eq:twchannelasymm}
&&\!\!\!\!\!A_{gb\rightarrow Wt}
=\frac{N(\mathbf{s}_\perp=1)-N(\mathbf{s}_\perp=-1)}{N(\mathbf{s}_\perp=1)+N(\mathbf{s}_\perp=-1)}\nonumber\\
&=&\!\!\!\!\!{\rm Im}C_{tW}\frac{v^2\sqrt{2s}m_W\lambda}{2\Lambda^2V_{tb}\left(
\sqrt{\lambda}((2m_W^2-3m_t^2)s-7(m_t^2+2m_W^2)(m_t^2-m_W^2))-2(m_t^2+2m_W^2)(\lambda+4 s m_t^2+(m_t^2-m_W^2)^2)\log y
\right)}\nonumber\\
&&\!\!\!\!\!-{\rm Im}C_{tG}\frac{2\sqrt{2}v m_t^2 s^{3/2}}{g_s\Lambda^2(s+m_t^2-m_W^2+\sqrt{\lambda})^3y^2}
\nonumber\\&&\!\!\!\!\!
\left\{
\left[(7m_t^2-8m_W^2)\lambda+4sm_t^2(11m_t^2-15m_W^2)-4m_t^2(m_t^2-m_W^2)^2\right]\left(\sqrt{\lambda}+s+m_t^2-m_W^2\right)
\right.\nonumber\\&&\!\!\!\!\!\left.
-8y\left[2(m_t^2-2m_W^2)(m_t^2-m_W^2)+s(3m_t^2-4m_W^2)\right]\left[\lambda+(s+m_t^2-m_W^2)\sqrt{\lambda}+2sm_t^2\right]
\right\}
\nonumber\\&&\!\!\!\!\!
\left(
\sqrt{\lambda}\left(s(2m_W^2-3m_t^2)-7(m_t^2+2m_W^2)(m_t^2-m_W^2)\right)-2(m_t^2+2m_W^2)(\lambda+4 s m_t^2+(m_t^2-m_W^2)^2)\log y
\right)^{-1}
\end{eqnarray}
where
\begin{equation}
\lambda=s^2+m_t^4+m_W^4-2sm_t^2-2sm_W^2-2m_t^2m_W^2
\end{equation}
and
\begin{equation}
y=\sqrt{\frac{s+m_t^2-m_W^2-\sqrt{\lambda}}{s+m_t^2-m_W^2+\sqrt{\lambda}}}
\end{equation}

In practice there is no way to measure the top spin directly, so we need to use the momentum of the decay products as the spin analyzer. The positron has a spin analyzing power $\alpha_{e^+}=1$. It can be shown that, if the top production process is followed by a semileptonic decay, one can replace the top spin in the triple-product $T$ by the positron three-momentum, and the corresponding asymmetry will be decreased by a factor of 1/2. For example, in the $s$-channel process, consider \begin{equation}
T=
(\mathbf{p}_u\times\mathbf{p}_{\bar{d}})\cdot\mathbf{p}_{e^+}
\end{equation}
We find
\begin{eqnarray}
A_{u\bar{d}\rightarrow t\bar{b}}&=&\frac{N(T>0)-N(T<0)}{N(T>0)+N(T<0)}\nonumber\\
&=&-{\rm Im}C_{tW}
\frac{3\pi v^2\sqrt{s}(s-m_t^2)}{4\sqrt{2}\Lambda^2V_{tb}m_W(2s+m_t^2)}
\end{eqnarray}
which is exactly half of Eq.~(\ref{eq:schannelasymm}), as expected. Similarly, for $t$-channel and $gb\rightarrow tW$ channel, the results in Eqs.~(\ref{eq:t1channelasymm}), (\ref{eq:t2channelasymm}) and (\ref{eq:twchannelasymm}) should also be reduced by a factor of $1/2$.
Note that although the CP-odd operator has effects on both production and decay processes, this asymmetry only reflects its effect on the production, because the decay process is only used as the spin analyzer, and the analyzing power $\alpha_{e^+}=1$ is not affected by the CP-odd effect.

We can also reverse the procedure and construct a CP-odd observable that only reflects the CP-odd effect in the decay process. In single top production, the top spin in its rest frame is always in the direction of the $d$ or $\bar{d}$ quark \cite{Mahlon:1996pn}. Although this gets modified by the operator $O_{tW}$ in the production vertex as is shown in Eqs.~(\ref{eq:topspin1}), (\ref{eq:topspin2}), the direction in which the top spin deviates is independent of the decay process, and thus the leading order effect gets averaged out as one considers the asymmetries. Therefore the $d$ or $\bar{d}$ three-momentum can be used to replace the top spin in Eq.~(\ref{T1}):
\begin{equation}
T=(\mathbf{p}_b\times\mathbf{p}_{e^+})\cdot \mathbf{p}_{d,\bar{d}}
\end{equation}
and the asymmetry becomes
\begin{eqnarray}
A_{t\to Wb}&=&\frac{N(T>0)-N(T<0)}{N(T>0)+N(T<0)}\nonumber\\
&=&{\rm Im}C_{tW}\frac{3\pi v^2(m_t^2-m_W^2)}{4\sqrt{2}\Lambda^2V_{tb}(m_t^2+2m_W^2)}
\end{eqnarray}
which agrees with Eq.~(\ref{eq:tdecayasymm}).

\subsection{CP-Violation in Top Pair Production}

The CP-violation effects in top pair production and decay have been considered in the literature before.
Refs.~\cite{Bernreuther:1993hq,Valencia:2005cx} have considered the CP-violation effect in the multi-Higgs doublet extensions of the SM. The effect of the top quark ``chromoelectric'' dipole moment, which corresponds the operator $O_{tG}$ with an imaginary coefficient, can be found in Refs.~\cite{Atwood:1992vj,Antipin:2008zx,Brandenburg:1992be}, where \cite{Brandenburg:1992be} has also considered the other two operators, $O_{\tilde{G}}$ and $O_{tW}$.
An analysis of the lepton transverse energy asymmetry at the Tevatron can be found in Ref.~\cite{Grzadkowski:1997yi}.
A recent numerical study of the ATLAS sensitivity to the complex phase of the $Wtb$ anomalous couplings can be found in Ref.~\cite{AguilarSaavedra:2007rs}.
The CP-violation effects of the top quark at linear colliders and photon colliders are discussed in
Refs.~\cite{linear1,linearCP,Grzadkowski:2003tf}.

In the top pair production processes, there are three operators that will contribute to CP violating observables:
\begin{eqnarray}
O_{tG}&=&(\bar{q}\sigma^{\mu\nu}\lambda^A t)\tilde{\phi}G^A_{\mu\nu}\nonumber\\
O_{\tilde{G}}&=&f_{ABC}\tilde{G}^{A\nu}_\mu G^{B\rho}_\nu G^{C\mu}_\rho\nonumber\\
O_{\phi\tilde{G}}&=&\frac{1}{2}(\phi^+\phi)\tilde{G}^A_{\mu\nu}G^{A\mu\nu}
\end{eqnarray}
where $\tilde{G}_{\mu\nu}=\epsilon_{\mu\nu\rho\sigma}G^{\rho\sigma}$. The first one contributes to both $gg\rightarrow t\bar{t}$ and $q\bar{q}\rightarrow t\bar{t}$ channels, while the last two contribute only to the $gg\rightarrow t\bar{t}$ channel.

A natural choice of the CP-odd observable is the triple-product considered in single top production. One could define similar quantities such as
\begin{equation}\label{eq:TPodd}
T=(\mathbf{p}_g\times\mathbf{p}_g)\cdot \mathbf{s}_t
\end{equation}
However this quantity doesn't result in any asymmetry, because the three CP-odd operators are P-odd but C-even. For both $gg\rightarrow t\bar{t}$ and $q\bar{q}\rightarrow t\bar{t}$ channels, under $PT_N$ symmetry the initial and final state do not change, except that the spins of $t$ and $\bar{t}$ are flipped. This means that $T$ defined in Eq.~(\ref{eq:TPodd}) is $PT_N$-odd and therefore the C-even operators cannot result in non-zero expectation values for $T$. We will need the spin information of both $t$ and $\bar{t}$ to observe CP-violation effect.

Here we define our CP-odd observables in a different way than the usual CP-odd triple product in most of the literature. In the top quark semileptonic decay, the amplitude contains a factor which is the inner product of the top spin and the lepton spin \cite{Atwood:1992vj}, and therefore we can use the spin projection operator to project the top spin on to the direction of the lepton three-momentum and ignore the other two decay products, in order to reduce the problem to a 2 to 2 scattering problem.

Consider the quark channel process $q\bar{q}\rightarrow t\bar{t}$ followed by the semileptonic decays of both $t$ and $\bar{t}$ quarks. We choose the coordinate axes such that in the CM frame, the top quark momentum is in the $z$-direction, the $q$ and $\bar{q}$ momenta are in $xz$ plane, and the angle between the $q$ and $t$ momenta is $\theta$. Let $\hat{\mathbf{p}}_{e^+}=(\sin\alpha_1\cos\beta_1,\sin\alpha_1\sin\beta_1,\cos\alpha_1)$ be the unit vector of the positron three-momentum in the top rest frame, $\hat{\mathbf{p}}_{e}=(\sin\alpha_2\cos\beta_2,\sin\alpha_2\sin\beta_2,\cos\alpha_2)$ be the unit vector of the electron three-momentum in the anti-top rest frame, and $\mathbf{v}=(\cos\theta,0,\sqrt{1-\beta^2}\sin\theta)$. Define the following triple-product:
\begin{equation}
T=
(\hat{\mathbf{p}}_{e^+}\times\hat{\mathbf{p}}_{e})\cdot\mathbf{v}
\end{equation}
we find that the the contribution from $O_{tG}$ can be written as:
\begin{equation}
\frac{d\sigma}{d\cos\theta d\cos\alpha_1 d\beta_1 d\cos\alpha_2 d\beta_2}
=-{\rm Im}C_{tG}\frac{g_s^3v\beta^2\sin\theta}{23328\sqrt{2}\pi^3\Lambda^2\sqrt{s}}T
\end{equation}
Clearly $T$ leads to an asymmetry:
\begin{eqnarray}
A_{q\bar{q}\rightarrow t\bar{t}}&=&\frac{N(T>0)-N(T<0)}{N(T>0)+N(T<0)}\nonumber\\
&=&-{\rm Im}C_{tG}\frac{\pi\sqrt{s}v\sqrt{1-\beta^2}}{2\sqrt{2}g_s\Lambda^2\beta(3-\beta^2)}
\left(K\left(\frac{\beta^2}{\beta^2-1}\right)-(1-2\beta^2)E\left(\frac{\beta^2}{\beta^2-1}\right)\right)
\end{eqnarray}
where
\begin{equation}
K(k^2)=\int^{\pi/2}_0\frac{d\theta}{\sqrt{1-k^2\sin^2\theta}}
\end{equation}
and
\begin{equation}
E(k^2)=\int^{\pi/2}_0 \sqrt{1-k^2\sin^2\theta}d\theta
\end{equation}
are the complete elliptic integral of the first and the second kind.
The SM has no contribution to this asymmetry because $T$ is parity-odd while the strong interaction is parity-even.

Now consider the gluon channel $gg\rightarrow t\bar{t}$. We use the same coordinate system, i.e. top quark momentum is in the $z$-direction and gluon momenta are in the $xz$ plane. $\hat{\mathbf{p}}_{e^+}$ ($\hat{\mathbf{p}}_{e}$) is the unit vector in the direction of the momentum of the positron (electron) in the top (anti-top) rest frame. Define two triple-products $T_z$ and $T_x$:
\begin{eqnarray}
T_z=(\hat{\mathbf{p}}_{e^+}\times\hat{\mathbf{p}}_{e})\cdot\mathbf{\hat{z}}\\
T_x=(\hat{\mathbf{p}}_{e^+}\times\hat{\mathbf{p}}_{e})\cdot\mathbf{\hat{x}}
\end{eqnarray}
The cross-section due to the CP-odd operators is
\begin{equation}
\frac{d\sigma}{d\cos\theta d\cos\alpha_1 d\beta_1 d\cos\alpha_2 d\beta_2}=
{\rm Im}C_{tG}(f_{tG}^z T_z+f_{tG}^x T_x)
+C_{\tilde{G}}(f_{\tilde{G}}^z T_z+f_{\tilde{G}}^x T_x)
+C_{\phi \tilde{G}}f_{\phi \tilde{G}}^z T_z
\end{equation}
where
\begin{eqnarray}
f_{tG}^z&=&-\frac{g_s^3v\beta^2}{248832\sqrt{2}\pi^3\Lambda^2\sqrt{s}(1-\beta^2\cos^2\theta)^2}\sqrt{1-\beta^2}
\nonumber\\&&
\left(9\beta^4\cos^6\theta+(7\beta^2-18\beta^4)\cos^4\theta
+(18\beta^4-25\beta^2+16)\cos^2\theta+7(2\beta^2-3)\right)
\\
f_{tG}^x&=&\frac{g_s^3v\beta^2}{248832\sqrt{2}\pi^3\Lambda^2\sqrt{s}(1-\beta^2\cos^2\theta)^2}
\nonumber\\&&
\left(9\beta^4\cos^4\theta+(7\beta^2-9\beta^4)\cos^2\theta-(23\beta^2-16)\right)\sin\theta\cos\theta
\\
f_{\tilde{G}}^z&=&\frac{3g_s^3\beta^2(1-\beta^2)\cos^2\theta}{165888\pi^3\Lambda^2(1-\beta^2\cos^2\theta)}
\\
f_{\tilde{G}}^x&=&-\frac{3g_s^3\beta^2\sqrt{1-\beta^2}\sin\theta\cos\theta}{165888\pi^3\Lambda^2(1-\beta^2\cos^2\theta)}
\\
f_{\phi\tilde{G}}^z&=&-\frac{g_s^2 s \beta^2(1-\beta^2)}{165888\pi^3\Lambda^2(s-m_h^2)(1-\beta^2\cos^2\theta)}
\end{eqnarray}
In general, any quantity that has the form $T(\mathbf{\hat{a}})=(\hat{\mathbf{p}}_{e^+}\times\hat{\mathbf{p}}_{e})\cdot\mathbf{\mathbf{\hat{a}}}$ may lead to an asymmetry. Using the following property of $T(\mathbf{\hat{a}})$:
\begin{equation}
\int\mbox{d}\Omega_{e^+}\mbox{d}\Omega_{e} T(\mathbf{\hat{a}})\mbox{sign}\left(T(\mathbf{\hat{b}})\right)=2\pi^3\left(\mathbf{\hat{a}}\cdot\mathbf{\hat{b}}\right)
\end{equation}
we find the asymmetry of $T(\mathbf{\hat{a}})$ is
\begin{eqnarray}
&&\frac{d\sigma(T(\mathbf{\hat{a}})>0)}{d\cos\theta}
-\frac{d\sigma(T(\mathbf{\hat{a}})<0)}{d\cos\theta}
\nonumber\\
&=&2\pi^3\left[
{\rm Im}C_{tG}(f_{tG}^z a_z+f_{tG}^x a_x)
+C_{\tilde{G}}(f_{\tilde{G}}^z a_z+f_{\tilde{G}}^x a_x)
+C_{\phi \tilde{G}}f_{\phi\tilde{G}}^z a_z
\right]
\end{eqnarray}

\section{Summary}

We have considered the effects of dimension-six operators in top quark physics.
The analysis is linear in the coefficients of these operators,
therefore the deviation from the SM is the interference terms between the SM and the new operators.
In general, integrating out heavy particles leads not to just one but to several
operators whose coefficients are related.
Therefore it is necessary to consider all dimension-six operators simultaneously.
Fortunately, although the total number of these operators is large,
we found that there are only 15 operators that can have significant interference terms.
In addition, for each decay or production process, only a few of them will contribute.
This is one of the advantages of the effective field theory approach:
although we don't have any knowledge of the new physics beyond the SM,
by making use of power counting and symmetries,
the number of parameters required to describe the new physics can be largely reduced.

We have obtained the deviation from the SM caused by these operators.
This allows us to constrain the new physics in a systematic way.
For example, we can measure (or put bounds on) the operator $O_{tW}$ by measuring
the $W$ boson helicity fraction $F_{L,R,0}$ and the analyzing power $\alpha_{b,\nu}$,
and then use the $s$- and $t$-channel single top production to put bounds on
$O_{\phi q}^{(3)}$ and $O_{qq}^{(1,3)}$.
The operator $O_{tG}$ can be constrained from the $Wt$ associated production
and the gluon channel $t\bar{t}$ production,
while the latter process also constrains $O_G$ and $O_{\phi G}$.
Finally, the quark channel $t\bar{t}$ production can be used to put bounds
on the four linear combinations of the four-quark operators.

The CP-violation effects in top quark physics are of particular interest. We have calculated the spin asymmetries caused by the 4 CP-odd operators. The observation of these asymmetries can be evidence of physics beyond the SM. In the single top production, these are the spin asymmetries in the direction perpendicular to the production plane. One could use the top decay process as a spin analyzer to study the asymmetry in the top production process, or vice versa. In $t\bar{t}$ production, we showed that both the top quark spin and anti-top quark spin are required to construct CP-odd observables.
\section*{Acknowledgments}

We are grateful for correspondence with Tim Tait and we thank C\'eline 
Degrande for checking many of our results.
This work was supported in part by the
U.~S.~Department of Energy under contract No.~DE-FG02-91ER40677.

\end{document}